\documentclass{article}
\usepackage[left=3cm,right=3cm,top=3cm,bottom=3cm]{geometry} 
\usepackage{amsmath} 
\usepackage{authblk}
\usepackage{amsmath, amssymb, enumerate, amsthm}

\usepackage{graphicx,psfrag,epsf}
\usepackage{graphicx,hyperref}
\usepackage{amssymb}
\usepackage{amsmath}
\usepackage{multirow}
\usepackage{float}
\usepackage{mathtools}
\usepackage{xcolor}
\mathtoolsset{showonlyrefs=true}
\usepackage{natbib}
\usepackage{subfig}
\usepackage{lipsum}
\usepackage{graphicx}
\usepackage{epsfig}
\usepackage{epstopdf}
\usepackage{xr}
\usepackage{fancyhdr,lipsum}
\DeclarePairedDelimiter\abs{\lvert}{\rvert}

\makeatletter

\newsavebox{\@linebox}
\savebox{\@linebox}[3em][t]{\parbox[t]{3em}{%
		\@tempcnta\@ne\relax
		\loop{\underline{\scriptsize\the\@tempcnta}}\\
		\advance\@tempcnta by \@ne\ifnum\@tempcnta<48\repeat}}

\usepackage{amssymb}

\begin{document}
	

\title{New formulation of the Logistic-Gaussian process to analyze trajectory tracking  data   }

\author{{Gianluca} {Mastrantonio}}
\affil{Politecnico di Torino, Dipartimento di Scienze Matematiche}
\author{ {Clara} {Grazian}}
\affil{Universit\`a degli Studi ``Gabriele d'Annunzio'', Dipartimento di economia \& University of New South Wales, School of Mathematics and Statistics}
\author{{Sara} {Mancinelli} }
\affil{Universit\`a di Roma ``La Sapienza'', Dipartimento di Biologia e Biotecnologie }
\author{{Enrico}  {Bibbona} }
\affil{Politecnico di Torino, Dipartimento di Scienze Matematiche}

	\date{}

	\maketitle

			\begin{abstract}

Improved communication systems, shrinking battery sizes and the price drop of tracking devices have led to an increasing availability of trajectory tracking  data. These data  
 are often analyzed to understand animal behavior.

In this work, we propose a new model for interpreting the animal movent as a mixture of characteristic patterns, that we interpret as different behaviors. The probability that the animal is behaving according to a specific pattern, at each time instant, is non-parametrically estimated using the Logistic-Gaussian process. Owing  to a new formalization and the way we specify the coregionalization matrix of the associated multivariate Gaussian process, our model is invariant with respect to the choice of the reference element and of the ordering of the probability vector components. We fit the model under a Bayesian framework, and show that the Markov chain Monte Carlo algorithm we propose is straightforward to implement.

We perform a simulation study with the aim of showing the ability of the estimation procedure to retrieve the model parameters. We also test the performance of the information criterion we used to select the number of behaviors. The model is then applied to a real dataset where a wolf has been  observed before and after procreation. The results are easy to interpret,  and  clear differences emerge in the two {phases}.

		\end{abstract}

		\maketitle

\section{Introduction}

Global Positioning System (GPS) telemetry is currently  the main tool used to  remotely determine  the  position of an animal with high precision, and at time intervals that can be programmed by the researcher \citep{Cagnacci2010}. These data, which are often defined as ``trajectory tracking  data'', take  the form of a time series of coordinates. The large amount of data, gathered from  GPS collars, facilitates a greater resolution in the study of habitat selection \citep{Hebblewhite}, spatio-temporal movements \citep{morales2004,Frair2010} and  behavior \citep{MERRILL2000,Anderson2003}.  For recent reviews,  the reader may refer to \cite{Patterson2017} and  \cite{hooten2017animal}

In this work, we propose a new statistical model for the analysis of trajectory tracking  data. The model can detect patterns in an animal's movement, that are  here interpreted  as consequence of  different behaviors.  The motivation for this study arises from a dataset that was collected to study the behavior of a female wolf, that was observed in the Abruzzo, Lazio and Molise National Park in the central Apennines, Italy \citep{Mancinelli2018}, before and after reproduction.

Animal movement modeling has a long history and, starting from the diffusion model of \cite{Brownlee1912}, a wide range of different approaches  have been proposed. These can be  grouped into three categories: \textit{i)}  point process models  \citep{Johnson2008b,Johnson2013,Brost2015}; \textit{ii)} continuous-time dynamic models \citep{BLACKWELL199787,Johnson2008a,Brillinger,fleming2014non}; \textit{iii)} discrete-time dynamic models \citep{morales2004,Jonsen2005,McClintock2012}. Point process  models require a numerical integration to compute the likelihood \citep{Cressie1993}  that  limit their use, while the  approaches based on  continuous-time dynamics  generally  fail to model  heterogeneity in the trajectory over time; notable exceptions can be found in \cite{hooten2017basis}, \cite{hooten2018animal} and \cite{blackwell2016exact}. 
Our proposal is part of the  discrete-time dynamic framework. 

In a discrete-time dynamic model, the joint distribution of the coordinates, or of their appropriate transformation, is  seen as a mixture process. Each component (or regime) represents a different movement pattern that may be interpreted as a particular kind of behavior. 
Switching between behaviors is often assumed to be temporally structured \citep{Houston1999,morales2004} and sometimes even spatially structured, as in  \cite{Blackwell2003},  often  ruled by a non-observed Markov process that leads to the class of  hidden Markov models  (HMMs)  in discrete (DT-HMM) \citep{zucchini2009b} and continuous-time (CT-HMM) \citep{blackwell2018integrated}. 

Although HMMs have been widely adopted  \cite[see, for example,][]{Franke2006,Langrock2012,Maruotti2015a}, they require 
knowledge or assumptions on the switching process between states, needed to define the rate functions; see, for example, \cite{morales2004}, \cite{patterson2009classifying} or \cite{blackwell2018integrated}, among others.  However, we believe that this  approach may be too restrictive to model animal trajectories. In this work, we  present a  discrete-time dynamic model, where the latent temporal  probability structure is estimated via  a non-parametric approach.  

We define a probability vector, also called compositional vector,  for each time. The elements of such vectors represent the probability that the animal is currently adopting a specific behavior. These vectors are modeled in a non-parametric fashion, which does not require prior knowledge of the possible temporal evolution, it can easily accommodate  covariate information,  it allows  dependence to be achieved between and within the probability vectors, and it is easy to implement.

We assume that the compositional vectors follow,   marginally, a Logistic-normal distribution ($\text{LogitN}$). 
The $\text{LogitN}$  has a representation in terms of normally distributed variables. Then, to introduce both serial dependence among the compositional vectors at different times, and internal dependence among the components of each vector, we  envision these variables as a realization of  a multivariate Gaussian process (GP) based on the coregionalization approach \citep{gelfand2010}. The  Logistic-Gaussian process ($\text{LogitGP}$) is then defined.

This is not the first construction that has used GPs  to introduce dependence over  compositional vectors. Nevertheless, our formalization  allows  inference to be performed in such a way that it enjoys two  properties, which we believe are essential in this setting:
i) invariance with respect to the choice of the reference element; ii) invariance with respect to the reordering of the labels that identify the behaviors. 
These two points have  {often} been overlooked in the literature \citep{brunsdon1998time,Paci2017} while, in 
 other cases, oversimplified hypotheses have been imposed and this has resulted in a reduced flexibility of the model \citep{Tjelmeland2003,Martins2016}.

In the recent literature, the flexibility of GPs has often  been  exploited  to model trajectory tracking data with different aims. We mention, for example, \cite{Johnson2008a,Fleming2,fleming2014non,calabrese2016ctmm, blackwell2016exact,Gurarie2017,scharf2018process,ctmm}. However, none of these references have any direct relationship with the methodology proposed here, since they use a GP to  directly model the observed locations. 

We  fit the model under a Bayesian framework, and propose a Markov chain Monte Carlo (MCMC) algorithm that is straightforward to implement. In order to fit the model, the number of latent behaviors must be specified,   and  an off-line procedure should then be used to select the best value. We suggest to use the integrated classification likelihood (ICL) \citep{Biernacki:2000} and we test its performance in a simulation study, and show that  it retrieves the right number of latent states most of the time. 

The model is applied to the dataset that motivated  the study. It consists of a series of GPS positions of a female wolf over two separate time windows, that correspond to different phases of the animal's life. In the first period, movements are more erratic, while  in the second, right after reproduction, movements are more regular, and
exhibit the classic star-shaped pattern around the den  \citep{Mech2003}.

Three behaviors are detected: one slow-speed behavior, which we call R (for Resting), and two high speed behaviors with different characteristics (HE1 and HE2, after Hunting/Exploring).
 Only R and HE1 are observed in the first time window, while  only R and HE2 are observed in the second time window. The temporal characterization of the R behavior is also different for the two time windows, since in  the first is more likely during daylight and in the second during the early hours of the night. A biological interpretation of these behaviors is investigated.

The paper is organized as follows. Section \ref{sec:data} presents the motivating example and a full description of the dataset that is analyzed. Section \ref{sec:model} describes the model, while  Section \ref{sec:logitn}  gives more details on the LogitGP, its properties and  its connections and differences from previous LogitGP-based models and CT-HMM. Section  \ref{sec:realdata} presents the results of the model fitted on the wolf data, and Section \ref{sec:conclu} concludes the paper.  We provide implementation details in Appendix  \ref{sec:impl}, while Appendix \ref{sec:simstud} presents the simulation study.

\section{Data}
\label{sec:data}

\begin{figure}[t]
    \centering
    {\subfloat[First time window]{\includegraphics[scale=0.35]{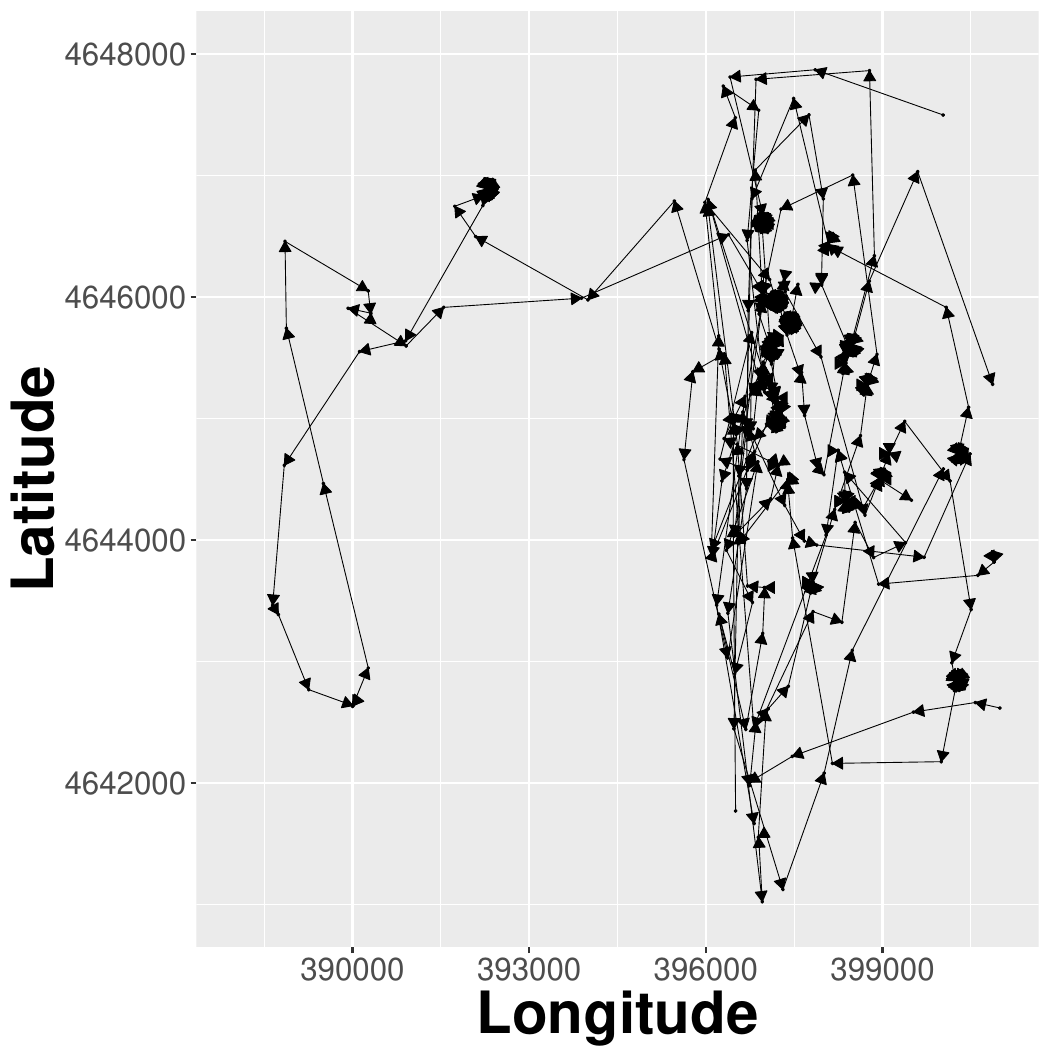}}}
    {\subfloat[Second time window]{\includegraphics[scale=0.35]{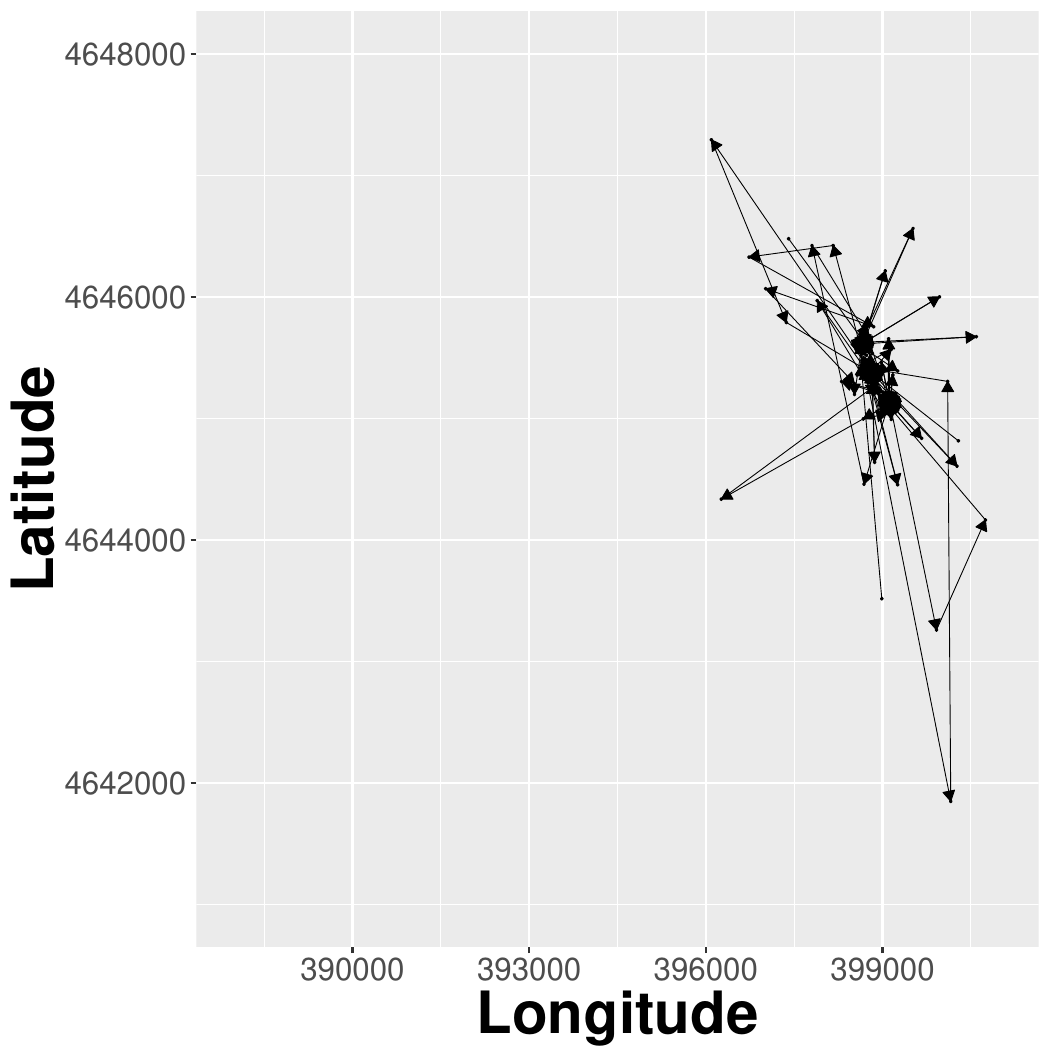}}}
    \caption{Trajectories in the two time windows}\label{fig:traj1}
\end{figure} 

\begin{figure}[t]
    \centering
    \includegraphics[scale=0.32]{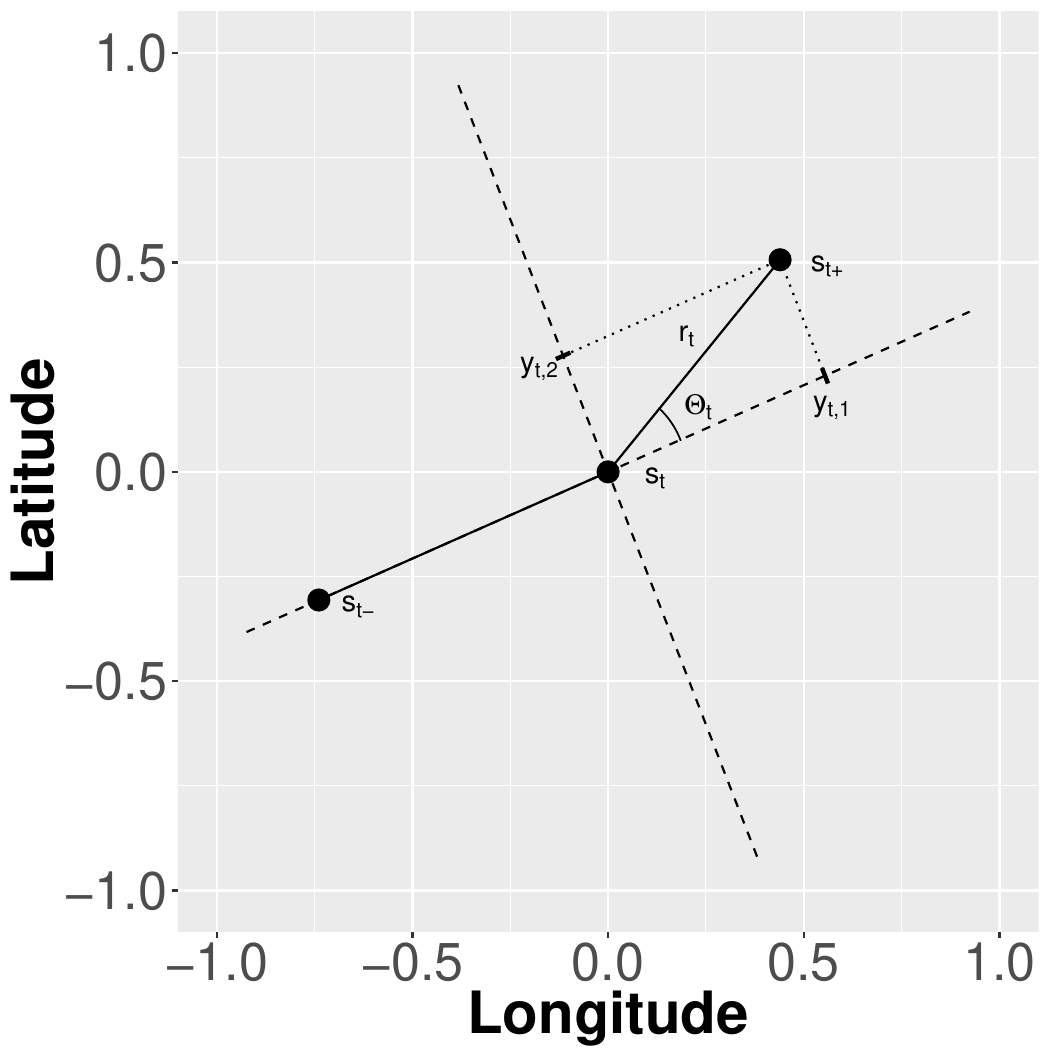}
    \caption{A graphical example of the relation between $\mathbf{s}_{t_i}$, ($\mathbf{r}_{t_i}, \theta_{t_i}$) and $\mathbf{y}_{t_i}$. }\label{fig:tras}
\end{figure} 
    
We have recorded a  time series of spatial locations  following the female wolf called F24 (2-3 years old), which was live-trapped in May 2009 in the Abruzzo, Lazio and Molise National Park (central Apennines, Italy) and equipped with a Vectronic Pro Light-1 collar (Vectronic Aerospace GmbH, Berlin, Germany). 
Details on the capturing  and handling of the wolf are provided in \cite{Mancinelli2018}.  During winter months (January –- April), the recording rate was scheduled  for every 30 minutes, 
while it was 
every 3 hours for the rest of the year. The higher acquisition rate during winter months was necessary to estimate wolf kill rates through field investigations of GPS clusters \citep{Sand2010}.

When first captured,  F24 was a member of the Villa pack, where it remained for 7.9 months before leaving and establishing a new pack (the Bisegna pack) in January 2010. The wolf gave birth in May 2010 and, using information derived from its GPS locations, it was possible to determine the actual position of the den. Between May 28 and June 4, F24 restricted its movements in the proximity to the den. Therefore, we assumed that F24 entered the den on May 28 and gave birth in the following days. Until collar failure occurred, on the 16th of June,  it was observed that F24 systematically revisited the den to feed and attend the cubs.  In order to detect the changes in behavior before and after wolf reproduction, we use a subset of the data relative to the final period in which F24 settled down with the Bisegna pack. The dataset is divided into two time windows, before and after reproduction,  which are described in Table  \ref{tab:obsper}. The trajectories are  shown in Figure \ref{fig:traj1}, where  it is possible to observe that the movements of F24 in the first period are more nomadic, while the classic  star-shaped pattern around the den, which characterizes the wolf reproductive period \citep{Mech2003}, can be observed in the second.
\begin{table}[t]
    \begin{tabular}{ccccc}
        \hline \hline
        Start & End & Interval &Number of obs. &Missing   \\
        \hline \hline
        03/01 00:00 &03/10 23:30 & 30 minutes & 473&7\\
        05/28 00:00 & 06/16 09:00 & 3 hours &  138 &18\\ \hline 
    \end{tabular}
    \caption{Observed time windows -  Temporal intervals, sample rates, number of observations and missing data.}\label{tab:obsper}
\end{table}

\subsection{Preliminaries and notation}
\label{sub:preli}

It should be noted  that the two time windows have different recording rates and there is a temporal gap between the two.
We denote $\mathcal{T}_j=(t^j_1,\cdots,t^j_{T_j} )$ the set of equally spaced time instants, separated by 30 minutes, which covers the $j$-th time window, and $\mathcal{T}=\mathcal{T}_1\cup\mathcal{T}_2$. The set of spatial locations is denoted by $\mathbf{s}= \{\mathbf{s}_{t}\}_{t \in \mathcal{T}}$, where each element $\mathbf{s}_{t}=(s_{t,1},s_{t,2})'$ is a  longitude and latitude vector.  The elements of $\mathbf{s}$ that are not observed are considered missing.

We define $t+$ to be the element that follows $t$ in $\mathcal{T}_j$, for every $t\in  \mathcal{T}_j$ but the last one. Similarly, we define $t-$ to be the element that precedes $t$ in $\mathcal{T}_j$,  for every $t\in  \mathcal{T}_j$ but the first one.   We define $\mathbf{y}_{t} = (y_{t,1},y_{t,2})'$, for $t\in  \mathcal{T}_j\backslash\{t^j_1,t^j_{T_j}\}$,  as the rotated vector of the coordinate increments, expressed in Km, between times $t$ and $t+$, so that $y_{t,1}$ is the projection of $\mathbf{s}_{t+}-\mathbf{s}_{t}$ onto the direction of the previous increment $\mathbf{s}_{t}-\mathbf{s}_{t-}$ (see Figure \ref{fig:tras}). The values of $\mathbf{y}_{t}$ with $t \in \{t^j_1,t^j_{T_j}\}$ are considered to be missing.
A classic approach for animal behavior modeling  involves the analysis of the following variables: the \emph{step-length} $r_{t} \in \mathbb{R}^{+}$, which is a proxy of the speed of the animal, and the \emph{turning-angle} $\theta_{t}\in [0,2\pi)$, which describes how the animal turns. These variables are transformations of  $\mathbf{y}_{t}$, as described in Figure \ref{fig:tras}. The explicit relation is $r_{t}=||\mathbf{y}_{t}||$ and $\theta_{t}=\mbox{atan}^* (y_{t,2},y_{t,1})$, where atan$^*$  is the two-arguments arctangent, which is a quadrant-specific inverse of the tangent function \citep{Merdia1999}.

The trajectory, within each time window, can then be described as
\begin{equation} 
\label{eq:traj}
\mathbf{s}_{t+} = \mathbf{s}_{t}+\mathbf{R}_{t-,t}\mathbf{y}_{t} \Rightarrow \mathbf{y}_{t} = \mathbf{R}_{t-,t}^{-1} \left( \mathbf{s}_{t+} - \mathbf{s}_{t}\right),
\end{equation}
where $\mathbf{R}_{t-,t}$ is the rotation matrix based on the angle $\theta_{t-}$. The variable $\mathbf{y}_{t}$ contains all the  information needed to describe the trajectory without losing any relevant  properties.

\section{The model}
\label{sec:model}


The coordinate increments $\mathbf{y}=\{\mathbf{y}_{t}\}_{t\in \mathcal{T}}$ are hypothesized to follow a mixture-type model with  conditionally independent data following a bivariate normal distribution, also called \emph{emission-distribution} in this context,  with parameters that depend on the mixture component; each component of the mixture represents a different kind of behavior. The clustering is encoded into a discrete (latent) random variable  $\mathbf{z}=\{z_t\}_{t\in \mathcal{T}}$, a membership variable that takes values in $\{ 1,2,\dots ,K \} \equiv \mathcal{K}$, where $z_t=k$ indicates that the behavior at time $t$ is the one identified as $k$. The latent variable $z_t$ is distributed according to a discrete distribution with probabilities $\boldsymbol{\pi}_t = \{\pi_{t,1}, \dots ,\pi_{t,K}\}$. 
These probabilities are realizations of a (latent) $\text{LogitGP}$, which is the main contribution of this paper. The latent $\boldsymbol{\pi}_t$  vary over time and are temporally correlated. Their distribution is described separately in Section \ref{sec:logitn}.
Schematically, the model is defined in the following way:
\begin{align}
f(\mathbf{y}|\mathbf{z} ,\{\boldsymbol{\xi}_k,\boldsymbol{\Omega}_k  \}_{k\in \mathcal{K}}) &= \prod_{t\in \mathcal{T}} \prod_{k \in \mathcal{K}}  \phi_2(\mathbf{y}_t|\boldsymbol{\xi}_{k},\boldsymbol{\Omega }_{k})^{\mathbb{I}_k(z_t)}, \label{eq:mod1} \\
z_t &\sim \text{Discrete}(\boldsymbol{\pi}_t), \label{eq:zetamodel} \\
\boldsymbol{\pi}(t) &\sim \text{LogitGP}(\mathbf{A},\boldsymbol{\mu}(t),\boldsymbol{C}(h)), \label{eq:pimodel}
\end{align}
where $\boldsymbol{\xi}_k$ and $\boldsymbol{\Omega}_k$ are the mean vector and covariance matrix of the bivariate normal density $ \phi_2(\cdot)$ for  component $k$, $\mathbb{I}_{b}(a)$ is the indicator function, which assumes a value of one when $a=b$, and zero otherwise. The probabilities $\boldsymbol{\pi}_t  = \{   \pi_{t,k} \}_{k \in \mathcal{K}}$ are discrete-time observations of an underlying and non-observed continuous-time process $\boldsymbol{\pi}(t)$, which is described   in Section \ref{sec:logitn}. For the moment, it is enough to say that $\boldsymbol{\pi}(t)$ is constructed from   a $K$-dimensional  GP  with a coregionalization matrix $\mathbf{A}$, a mean function $\boldsymbol{\mu}(t)$ and a vector of correlation functions $\boldsymbol{C}(h)$, where $h$ is a temporal distance.

In practice, given the latent variables ${z}_t$ (the behavior), the observations $\mathbf{y}_t$ are independent and normally distributed, with parameters  that depend on the current behavior (${z}_t$). 
The model described in \eqref{eq:mod1} assumes that all the observations ($\mathbf{y}_t$) follow the same distribution, which is only reasonable  if the temporal distances between consecutive observations are fixed. This is the reason why we consider a time grid with fixed increments, as stated in Section \ref{sec:data}.  
The fixed time difference between  the element of $\mathcal{T}$ forces us to estimate a large number of missing observations. Although this increases the computational time, it is necessary to model the two time windows together and compare them.

It should be noted that a bivariate normal distribution on $\mathbf{y}_t$ induces a projected normal  distribution on the turning-angle \citep{Wang2013}, which is one of the most flexible distributions for circular data \citep[see, for example,][]{mastrantonio2015b,mastrantonio2015}.
Unfortunately, no closed form is available for the  step-length.

\section{$\text{LogitN}$ distribution    and  $\text{LogitGP}$} 
\label{sec:logitn}

\cite{Aitchison1986} proposed the $\text{LogitN}$ distribution to model compositional data as an alternative to the Dirichlet distribution.  Although it has a closed form, we prefer to introduce it by using its constructive definition.

The vector $\boldsymbol{\pi}_t$ is defined as 
\begin{align} 
\pi_{t,k} = \frac{e^{\omega_{t,k}}}{\sum_{{j=1}}^Ke^{\omega_{t,j}}}, \quad  k \in 1,\dots , K \label{eq:piAA}
\end{align}
where $\omega_{t,k}$ are real valued variables.  It should be noted that adding a constant to each $\omega_{t,k}$ produces the same vector of probabilities, and  an identifiability constraint is therefore needed; without loss of generality, the $K-$th element  is set to  zero ($\omega_{t,K}=0$) and  is treated as the \emph{reference element}.
If $\boldsymbol{\omega}_t = \{ \omega_{t,k} \}_{k=1}^{K-1}$  is normally distributed, then $\boldsymbol{\pi}_t$ is said to be $\text{LogitN}$ distributed \citep{Aitchison1986}. It is possible to introduce dependence between the compositional vectors, while preserving the $\text{LogitN}$ as the marginal distribution, by envisioning  $\boldsymbol{\omega}_t $ as a realization of a $K-1$ dimensional  GP $\boldsymbol{\omega} (t)$.
  
Several proposals, that directly  define all the elements of $\omega_{t}$   as a  $(K-1)$-variate GP exist \cite[see, for example,][]{Paci2017,Tjelmeland2003,Martins2016,Pawlowsky1992}. Unlike those proposals,  we introduce an auxiliary $K$-dimensional  GP $\boldsymbol{\gamma} (t)$, with coregionalization matrix $\mathbf{A}$ which we require to be non-negative definite and symmetric, with  mean function  $\boldsymbol{\mu}(t)$ and a vector of correlation functions $\boldsymbol{C}(h)$. From $\boldsymbol{\gamma} (t)$, we construct $\boldsymbol{\omega}(t)$ as
\begin{align}
\omega_k(t) &= {\gamma_k}(t)-{\gamma_K}(t), \label{eqomegamodel}\\
\boldsymbol{\gamma}(t) & = \boldsymbol{\mu}(t)+\mathbf{A}\boldsymbol{\gamma}^*(t), \label{eq:gammamodel}\\ 
{\gamma}_k^*(t) & \sim \text{GP}({0}, C_k(h)).\label{eq:gammamodel2}
\end{align}
Matrix  $\mathbf{A}$ introduces dependence between the elements of $\boldsymbol{\gamma}(t),$ since $\boldsymbol{\Sigma}=\mathbf{A}\mathbf{A}'$ is the covariance of $\boldsymbol{\gamma}(t)$. The explicit relation between $\mathbf{A}$ and $\boldsymbol{\Sigma}$ is 
\begin{equation}
\mathbf{A}= \boldsymbol{\Delta}\boldsymbol{\Xi}^{\frac{1}{2}}\boldsymbol{\Delta}', \label{eq:A}
\end{equation}
where $\boldsymbol{\Delta}$ and $\boldsymbol{\Xi}$ are the matrix of the eigenvectors of $\boldsymbol{\Sigma}$ and the diagonal matrix of the eigenvalues, respectively. The above definition makes $\mathbf{A}$ the unique symmetric and non-negative square root of $\boldsymbol{\Sigma}$.
By stating that
\begin{equation}\boldsymbol{\pi}(t) \sim \text{LogitGP}(\mathbf{A},\boldsymbol{\mu}(t),\boldsymbol{C}(h)),\label{eq:piLGP}\end{equation}
as in equation \eqref{eq:pimodel}, we mean that  $\boldsymbol{\pi}(t)$ is constructed according to equations \eqref{eq:piAA}--\eqref{eq:A}.
We use this approach to avoid specifying a predefined functional form of $\pi(t)$ and to estimate $\pi(t)$ non-parametrically, using a GP, as  is generally done for Bayesian non-parametric inference; see, for example, \cite{muller2017bayesian}.

Equations \eqref{eq:piAA}-\eqref{eq:gammamodel2} introduce a serial correlation between vectors $\boldsymbol{\pi}_t$, which is required  if the aim is to make a behavior persist over time. It should be noted that $\boldsymbol{\omega} (t)$ is a linear combination of GPs and  it is therefore  a GP itself. 

Equation \eqref{eqomegamodel} implicitly defines the reference element as the $K-$th since, by construction,  $\omega_K(t) = {\gamma_K}(t)-{\gamma_K}(t)= 0$. 
From \eqref{eqomegamodel}, it is also possible to see that  the mean function $\boldsymbol{\mu}(t)$ is not identifiable, unless an identification constraint is introduced,  and  we set the component of $\boldsymbol{\mu}(t)$ relative to the reference element to zero:  $\mu_{K}(t)=0$.

It is important to highlight that $\boldsymbol{\gamma}(t)$ is not identifiable and any inference about $\boldsymbol{\pi}_t$ is in fact made by looking at $\boldsymbol{\omega}_t$ through equation \eqref{eq:piAA}.
The advantage of introducing $\boldsymbol{\gamma}(t)$ can be seen by replacing \eqref{eqomegamodel} in \eqref{eq:piAA}:
\begin{align} 
\pi_{t,k} = \frac{e^{\gamma_{t,k} -\gamma_{t,K} }}{\sum_{{j=1}}^Ke^{\gamma_{t,j}-\gamma_{t,K} }} = \frac{e^{\gamma_{t,k}}}{\sum_{{j=1}}^Ke^{\gamma_{t,j}}}, \quad k \in 1, \dots , K. \label{eq:piAA3}
\end{align}
With respect to \eqref{eq:piAA}, where $\omega_{t,K}$ must be set to zero, the right hand side of \eqref{eq:piAA3} has a more symmetric form, since all the components of $\boldsymbol{\pi}_{t}$ are written in terms of exponentials of $\gamma_k (t)$ and there is no reference element.

\subsection{Invariances} \label{sec:inv}

A desirable property of a general model  is that it should not depend on any arbitrary choice. In particular, it is desirable to have:
\begin{itemize}
	\item invariance from the choice of the reference element;
	\item invariance with respect to the reordering of the labels.
\end{itemize}
In order to demonstrate the first property, it is sufficient to observe that the right hand side of equation \eqref{eq:piAA3} does not depend on the choice of the reference element.

The second property means that  there exists a reparameterization of the model that maintains the likelihood invariant with respect to the reordering of the labels. It is well known that, when  labels in $\boldsymbol{\pi}(t)$ are permuted, the likelihood  function  remains invariant, if the parameters are permuted accordingly. It is therefore sufficient to show that, when \eqref{eq:piLGP} holds and if $\mathbf{P}$ is  a permutation matrix, 
\[\mathbf{P}\boldsymbol{\pi}(t) \sim \text{LogitGP}(\mathbf{P}\mathbf{A}\mathbf{P}',\mathbf{P}\boldsymbol{\mu}(t),\mathbf{P}\boldsymbol{C}(h)).\]
In order to understand this, let us consider the vector of the independent GPs $\tilde{\boldsymbol{\gamma}}^*(t)$ with  the correlation vector $\mathbf{P}\boldsymbol{C}(h)$. The new vector $\tilde{\boldsymbol{\gamma}}^*(t)$ has the same distribution as the reordered vector $\mathbf{P}{\boldsymbol{\gamma}}^* (t)$ from equation \eqref{eq:gammamodel}.
Moreover, if we apply the coregionalization matrix $\tilde{\mathbf{A}}=\mathbf{P}\mathbf{A}\mathbf{P}'$ to $\tilde{\boldsymbol{\gamma}}^*(t)$ and add the vector of the reordered means $\mathbf{P}\boldsymbol{\mu}(t)$, we obtain $\tilde{\boldsymbol{\gamma}}=\mathbf{P}\boldsymbol{\mu}(t)+\tilde{\mathbf{A}}\tilde{\boldsymbol{\gamma}}^*(t)$, which has the same distribution as $\mathbf{P}\boldsymbol{\gamma}(t)$. It should be pointed out  that the covariance matrix of $\mathbf{P}\boldsymbol{\gamma}(t)$ is $\tilde{\boldsymbol{\Sigma}}=\mathbf{P}\boldsymbol{\Sigma}\mathbf{P}'=\tilde{\mathbf{A}}\tilde{\mathbf{A}}'$, and that $\tilde{\mathbf{A}}$ is symmetric and non-negative definite.
In short, if we construct $\tilde {\boldsymbol{\pi}}(t)$ from $\mathbf{P}\boldsymbol{\gamma}(t)$ according to equation \eqref{eq:piAA3}, we obtain $\mathbf{P}\boldsymbol{\pi}(t)$.

Finally, it should also be pointed out that the standard choice for a coregionalization matrix in geostatistics, that is the Cholesky decomposition, suffers from some drawbacks in this setting: it is in fact well know that it introduces an ordering between the processes \citep{rothman2010new}. The triangular structure of  coregionalization matrix $\mathbf{A}$ is not preserved under a (non-diagonal) permutation matrix $\mathbf{P}$: if $\mathbf{A}$ is triangular, $\mathbf{P}\mathbf{A}\mathbf{P'}$ is  no longer  triangular, and it cannot therefore be the coregionalization matrix of the process $\mathbf{P}\boldsymbol{\gamma} (t)$.

\subsection{The log-ratio mean and variance} \label{sec:logratio}

The nature of compositional vectors makes the interpretability of the moments difficult. In fact, the sum-to-one constraint  restricts the domain to the simplex and induces negative correlations among the variables \cite[see][]{Aitchison1986}. More consistent definitions of the moments, based on the log-ratios between the components, were proposed in \cite{Aitchison1986}. These definitions are particularly easy to compute under our model specification: 
\begin{align}
\rho_{i,k}(t,t') =&\mathbb{E}\left(\log  \frac{\pi_{t,i}}{\pi_{t',k}} \right)= \mathbb{E}\left(\gamma_{t,i}- \gamma_{t',k}\right),\label{eq:rho1} \\
\tau_{ij,k}(t,t') =&  \mathbb{C}\mbox{ov}\left(\log  \frac{\pi_{t,i}}{\pi_{t,k}} , \log  \frac{\pi_{t',j}}{\pi_{t',k}}\right)  =\label{eq:tau1} \\
&\mathbb{C}\mbox{ov}\left(\gamma_{t,i}- \gamma_{t,k},\gamma_{t',j}-\gamma_{t',k}\right). \nonumber 
\end{align}
Let ${A}_{ij}$ be the $(i,j)$-th element of matrix $\mathbf{A}$, it is then possible to write \eqref{eq:rho1} and  \eqref{eq:tau1} in closed form:
\begin{align}
\rho_{i,k}(t,t')=&\mu_{t,i}-\mu_{t',k}, \label{eq:rho3} \\
\tau_{ij,k}(t,t') =&A_{ii} C_i(\abs{t'-t})(A_{ji}+A_{ki}) + \label{eq:tau3} \\
& A_{jj}C_j(\abs{t'-t})(A_{ij}+A_{kj}) + \nonumber\\
&  A_{kk}C_k(\abs{t'-t}) ( 2A_{kk}-A_{ik}-A_{jk})  \nonumber. 
\end{align}

Important properties of the compositional vectors $\boldsymbol{\pi}(t)$ can be established by imposing analogous properties on  GP $\boldsymbol{\gamma}(t)$.
For example, if the components of $\boldsymbol{\gamma}(t)$ are independent ($\mathbf{A}$ is diagonal), then the components of $\boldsymbol{\pi}(t)$ are subcompositionally independent \cite[see][Property 10.3]{Aitchison1986}, and the log-ratios are uncorrelated \cite[see][Section 5.9]{Aitchison1986}, since $\tau_{ij,K}(t,t)$ takes the particular expression
\[
\tau_{ij,K}(t,t)=\begin{cases} 2A_{ii}^2+2A_{KK}^2&\text{for } i=j \text{ and } i \neq K,\\
2A_{KK}^2&\text{for } i\neq j \text{ and } i,j\neq K. \end{cases}
\]
Moreover, if the components of  vector $\boldsymbol{\gamma}(t)$ are independent and identically distributed, we obtain $A_{ii}=A_{jj}$ and  $\rho_{i,j}(t,t)=0$ for $i,j=1,\dots , K$, which is the requirement of having $\boldsymbol{\pi}(t)$ subcompositionally independent and with elements identically distributed \citep{Aitchison1986}. The latter property can be seen immediately from  \eqref{eq:piAA3}.

\subsection{Relationship with other definitions of the \emph{LogitGP}}\label{sec:other}
A Bayesian non-parametric approach, based on a GP, was also  adopted in previous works. Most of the the existing constructions are derived from equation \eqref{eq:piAA}, where a GP has been used to directly model $\omega_{k}(t)$. As already mentioned, identifiability requires that a reference element, here the $K$-th, is chosen. The corresponding component $\omega_{t,K}$ is set to zero, and  its definition is hence deterministic and the covariance function of the reference element  therefore vanishes.
Unless special care is taken, changing the reference element can affect the structure of the log-ratio covariance of  $\pi(t)$, measured by \eqref{eq:tau3}. In fact,  if any of the indices $i,j,k$ is equal to $K$ in equation \eqref{eq:tau3}, only two correlation functions appear in the expression instead of three. To compensate for this effect, and to keep the expression of $\tau_{ij,k}$ invariant, it would be necessary to correspondingly adjust the other correlation functions in a non-trivial way, or to make some other ad-hoc assumptions. 
Our method provides a way of keeping this invariance in a natural way and in a very general framework.

In \cite{Paci2017}, the GPs $\omega_{k}(t)$ with $k\in \{1\cdots K-1\}$, are assumed to be independent. However a different choice of the reference element   is equivalent to using another set of GP $\boldsymbol{\omega}'(t)=\mathbf{L} \boldsymbol{\omega}(t)$ for a suitable non-diagonal matrix $\mathbf{L}$ \citep{Aitchison1986}. The new processes $\omega'_{k}(t)$ are correlated and no longer satisfy the same assumptions.
On the other hand, in \cite{Tjelmeland2003} and in \cite{Martins2016}, the problem is circumvented by additionally assuming that there is a single correlation function, that is common to all the components. This makes the model invariant, but at the price of introducing a restrictive constraint. Only GPs of the VARMA family are considered in \cite{brunsdon1998time}. Changing the reference element is again equivalent to introducing the transformation $\boldsymbol{\omega}'(t)=\mathbf{L}  \boldsymbol{\omega}(t)$ mentioned above. Such a transformation preserves the fact that the new processes are in the VARMA family. Nevertheless it does not preserve the structure of the log-ratio covariances and, therefore, it does not make the resulting inference about the $\pi(t)$ invariant.
A completely different proposal, which is not based on a coregionalization but on the co-kriging approach, can be found in \cite{Pawlowsky1992} and in \cite{Lark2007}. 

\subsection{Differences  from and similarities with the  CT-HMM} \label{sec:HMM}

One of the main competitors of the here proposed  model  is the CT-HMM  of \cite{blackwell2016exact}. 
However, the two approaches are somewhat different. 
To make the comparison straightforward, we still consider equation \eqref{eq:mod1} valid, but interpret the membership variable $z_t$ as a discretely observed continuous-time Markov chain $z(t)$, whose transition rates is $\lambda_{ij}(t)$.
In \cite{blackwell2016exact}, the number of behaviors and the transition rates are established in advance, and are considered as part of the model. The transition rates between behaviors depend on some parameters that are  inferred from data. For example, a simplified version of the wild boar example in \cite{blackwell2016exact} assumes that three behaviors exist: resting, foraging and returning. The animal  repeats these behaviors each day in the given fixed sequence. There exists a preferred time for each transition, making the Markov chain non-homogeneous. The transition rates follow a parametric function of time that needs to be specified in advance. We are in the framework of parametric inference.
 
On the other hand, in our approach, the framework is non-parametric and there is one more level in the hierarchical structure. Given a realization of $\pi(t)$, the $z_t$ are independent of  each other.  Since we do not want to give a predefined functional form to $\pi(t)$, we interpret it as a realization of an unobserved LogitGP, in analogy with what is done, for example, in non-parametric regressions \citep[see][Chapter 4.3.3]{muller2017bayesian}. We then let Bayesian inference reveal the a posteriori distribution of $\pi(t)$.

\section{Wolf data}
\label{sec:realdata}

We now present the application of the proposed method to the dataset described in Section \ref{sec:data}.  

In order to fully specify the model, we assume that each $\mu_k$ is a piecewise constant function whose value is $\mu_k(t)=\beta_{k,1}+ \chi_2(t)\beta_{k,2}$, where $\chi_2$ is a function that takes the value 0 if $t$ belongs to the first time window and 1 otherwise. Moreover, we assume that the correlation functions $C_k$ are all exponential: $C_k(h)=\text{e}^{-\psi_k \abs{h}}$. Under this assumption, the GP is Markovian and this makes computation much faster. Other  choices are possible,  but since the GP has to be evaluated over a large number of temporal points,  a geostatistical approximation is needed to avoid the so-called ``big N problem'' \citep{Jona2013b}. For a general  review of the methods that can be exploited,  see \cite{Heaton2017}.

The model parameters   are therefore $\left( \boldsymbol{\Sigma},\{\beta_{k,1}, \beta_{k,2}  \}_{k=1}^{K-1},\{\boldsymbol{\xi}_k,\boldsymbol{\Omega}_k, \psi_k  \}_{k= 1}^ K \right)$, where $\boldsymbol{\Sigma}= \mathbf{A}\mathbf{A}'$, and   a prior distribution must be specified for each of them. We assume independence between the parameters:   $\boldsymbol{\xi}_k \sim N_2(\mathbf{0}_2, 100\mathbf{I}_2)$ and $\boldsymbol{\Omega}_k \sim IW(3, \mathbf{I}_2)$  are used for the likelihood parameters,  $ \psi_k \sim U(0.3,6)$ for the temporal decays, $\beta_{k,j} \sim N(0,100)$, $j=1,2$, for the regression coefficients,  and  $\boldsymbol{\Sigma} \sim IW (K+1, \mathbf{I}_K)$  for the variance parameter of the LogitN process, where $IW(b,\mathbf{B})$ stands for the inverse Wishart distribution with $b$ degrees of freedom and scale matrix $\mathbf{B}$.

We compare our proposal  with models based on the same likelihood but with  different dynamics for $z_t$, in other words with a  DT-HMM and a  CT-HMM.  
We follow \cite{blackwell2016exact} for the  CT-HMM, and  the rate function from state $i$ to $j$ is defined as follows:
$$
\lambda_{i,j}=\frac{v_{i,j}}{1+e^{ v_{i,j}(t\text{ mod }1-t_{i,j}^0) }}, 
$$
while  the probability of switching from the $i-$th state to the  $j-$th, in the DT-HMM,  is given by $\pi_{i,j}$.
As prior distributions, we use  a  $U(0,1)$ for all the $t_{i,j}^0$, while,  following \cite{blackwell2018integrated}, we set  $v_{kj}/48 \sim Beta(1,1)$, where 48 is the number of time points in a day. The vector $(\pi_{i,1},\dots \pi_{i,K})'$  follows a Dirichlet distribution, with all the parameters equal to 1.

We test  $K \in \{2,\dots,6\} $ for all the models. 
The choice of the number of components is essential in applications involving mixtures. 
Since the models are defined via a latent variable ${z}_t$, we use the ICL, proposed by \cite{Biernacki:2000}. 
For a thorough review of the methods that can be used to perform model selection,  the reader may refer  to \cite{pohle2017selecting}.
The MCMC is implemented with 1,000,000 iterations, burnin 70,000 and thinning equal to 6, which results in 5,000 posterior samples. The models are implemented on the Bari ReCaS Data Center \citep{Farm}.

\subsection{The results}  

\begin{figure}[t!]
    \centering
    {\subfloat[First time window]{\includegraphics[scale=0.22]{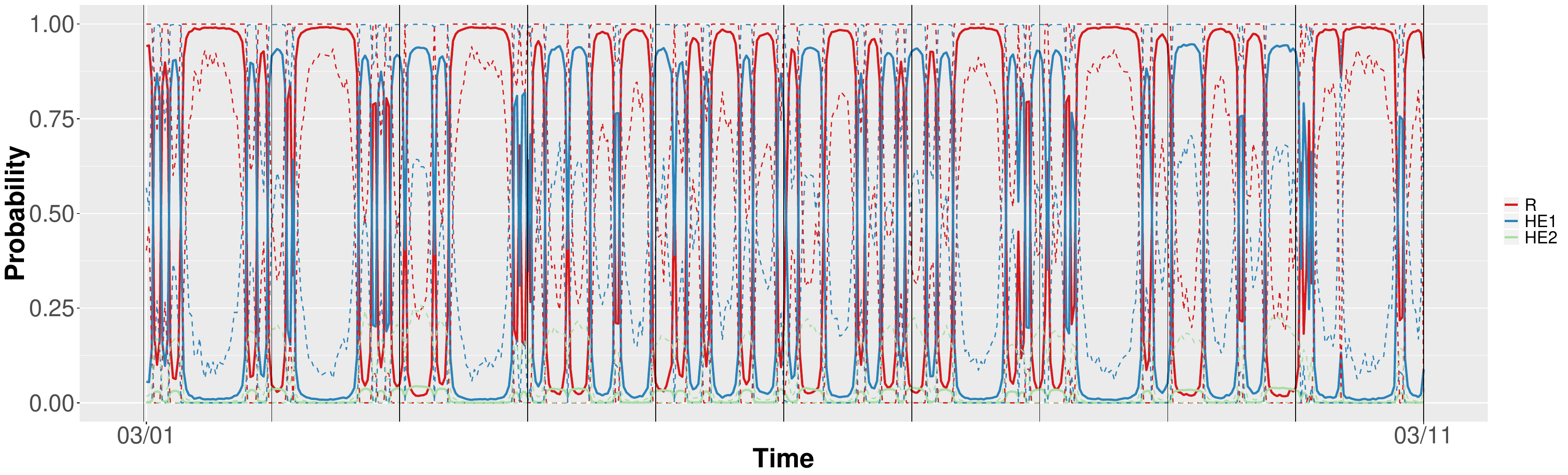}}}\\
   {\subfloat[Second time window]{\includegraphics[scale=0.22]{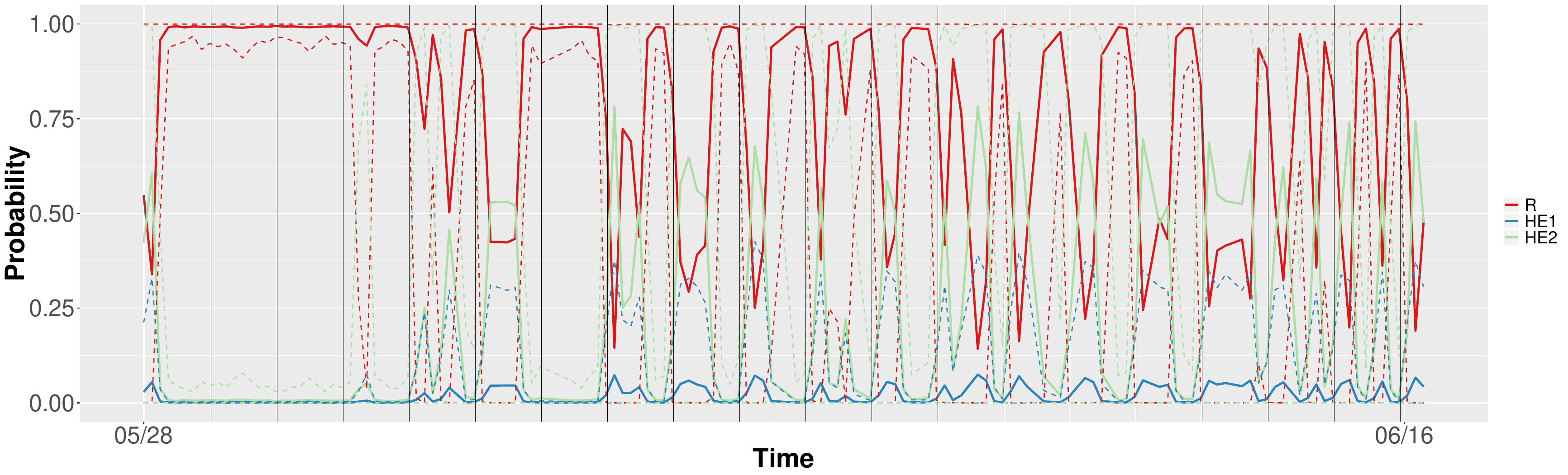}}}\\
    \caption{Posterior estimates of the probability vector time series. The solid lines represent the posterior means, while the dotted lines are the limits of the 95\% CIs. The vertical lines indicate 00:00 hours.
    } \label{fig:estTOT}
\end{figure} 
 
\begin{figure}[t!]
	\centering
	{\subfloat[]{\includegraphics[scale=0.25]{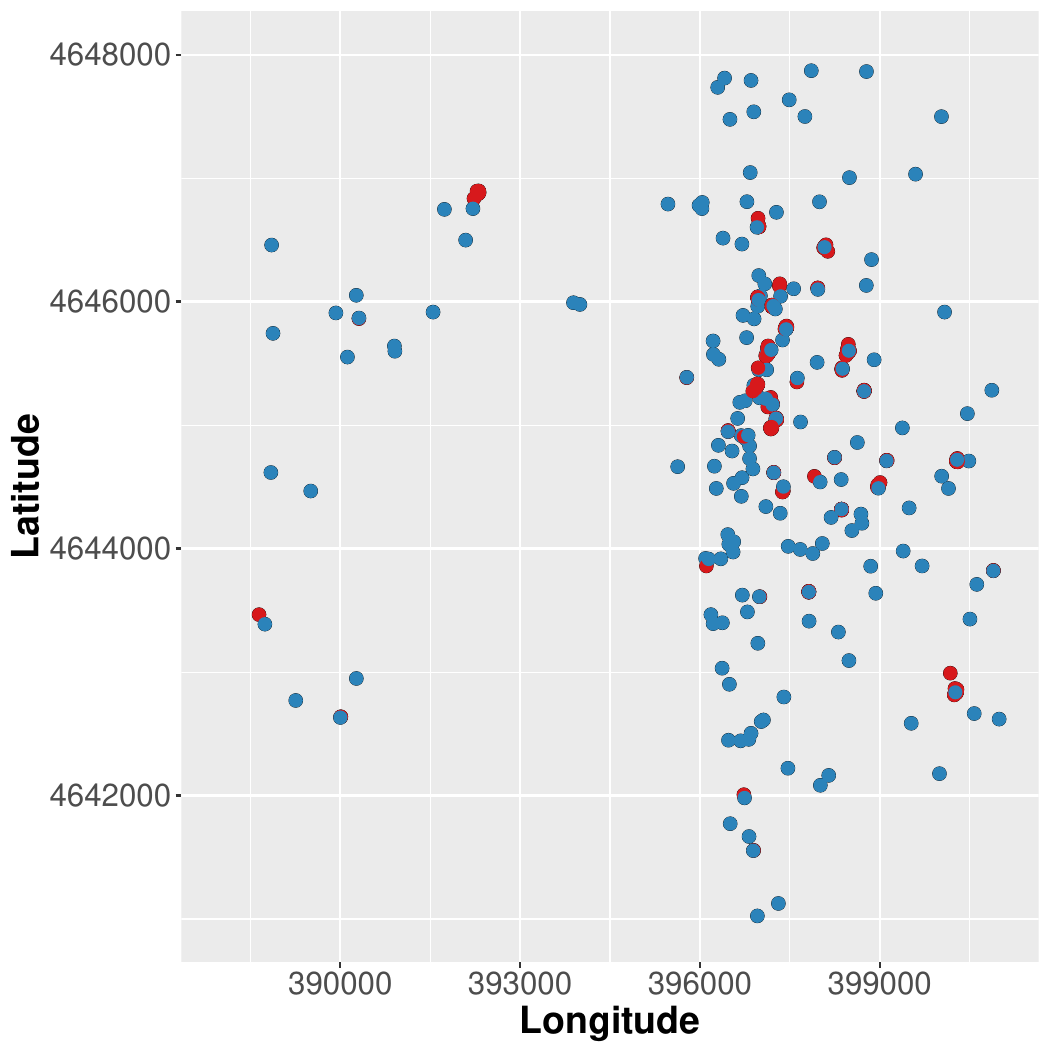}}}
    {\subfloat[]{\includegraphics[scale=0.25]{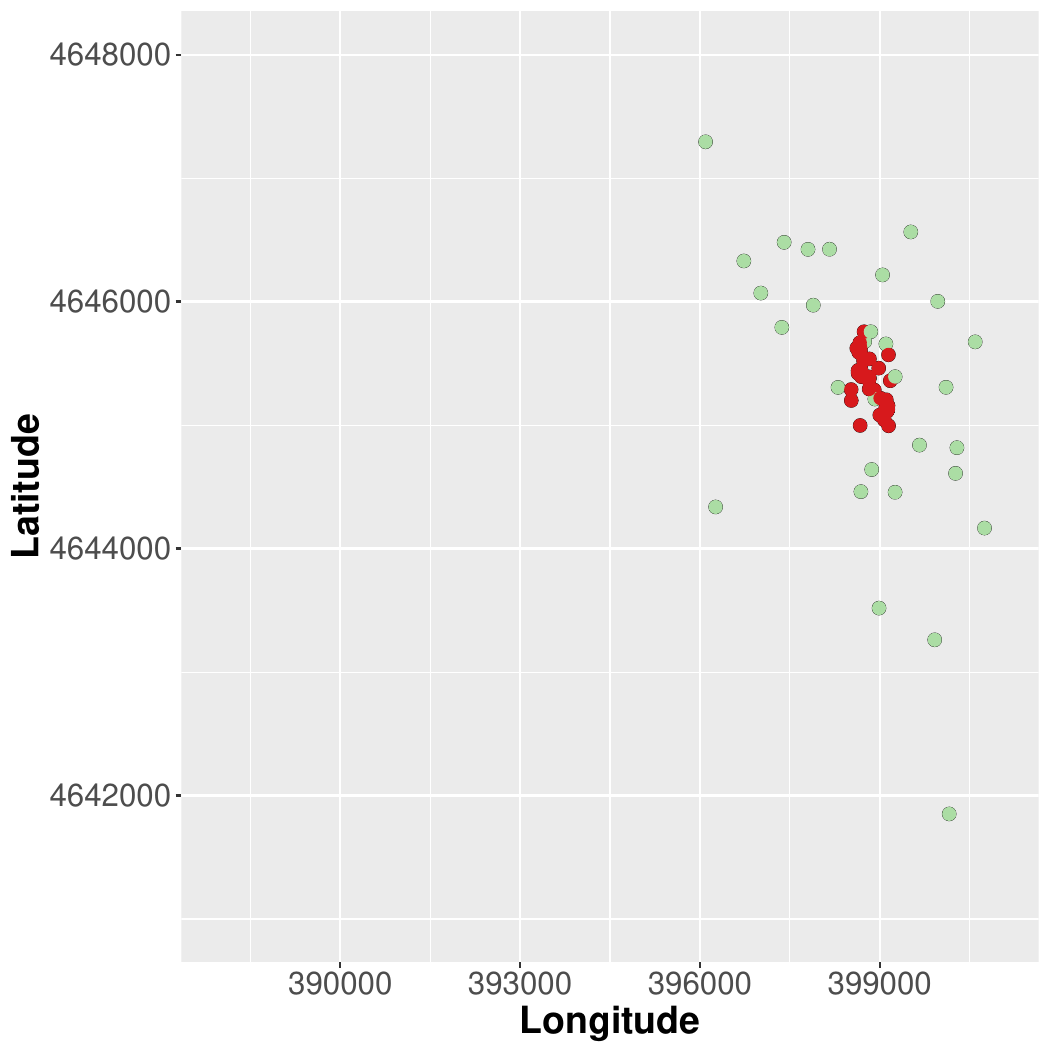}}}\\
	{\subfloat[]{\includegraphics[scale=0.25]{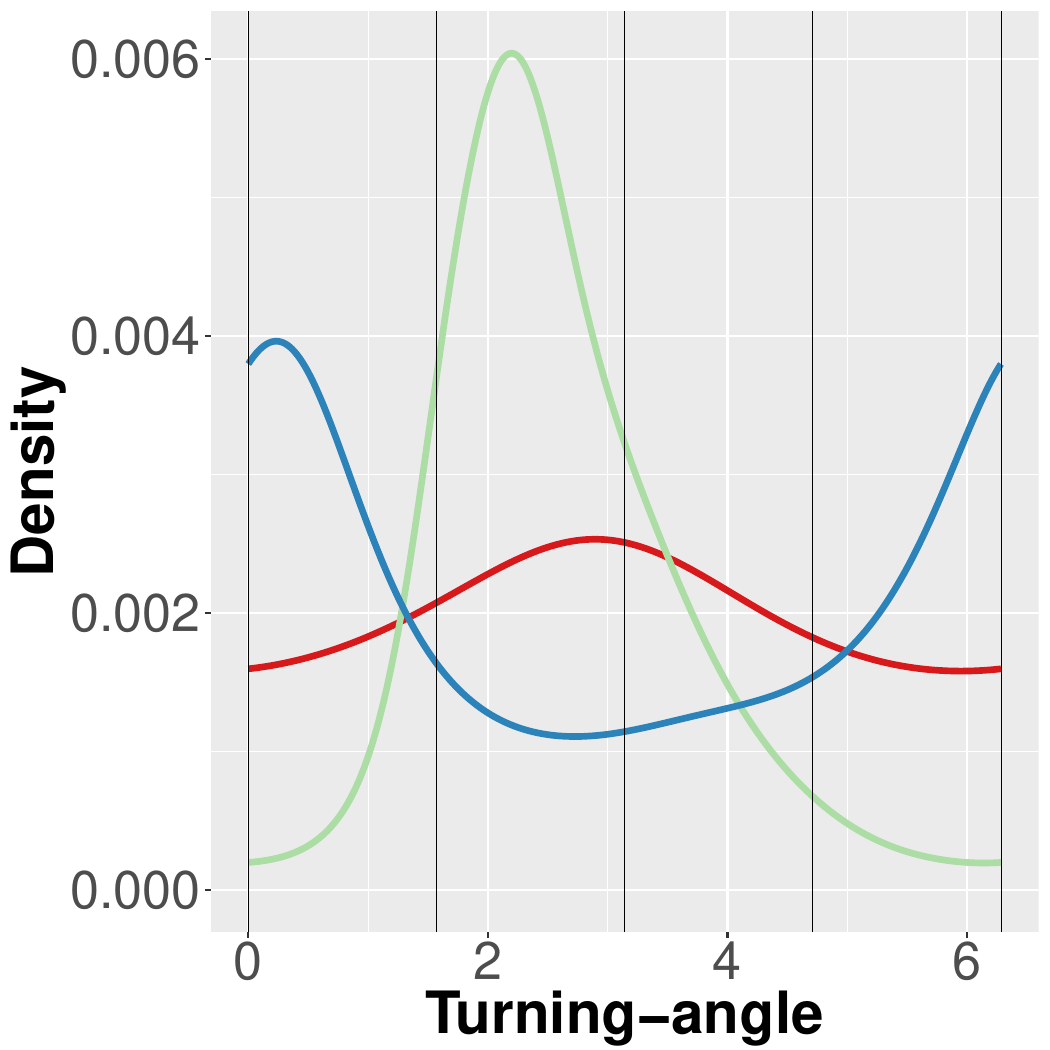}}}
	{\subfloat[]{\includegraphics[scale=0.25]{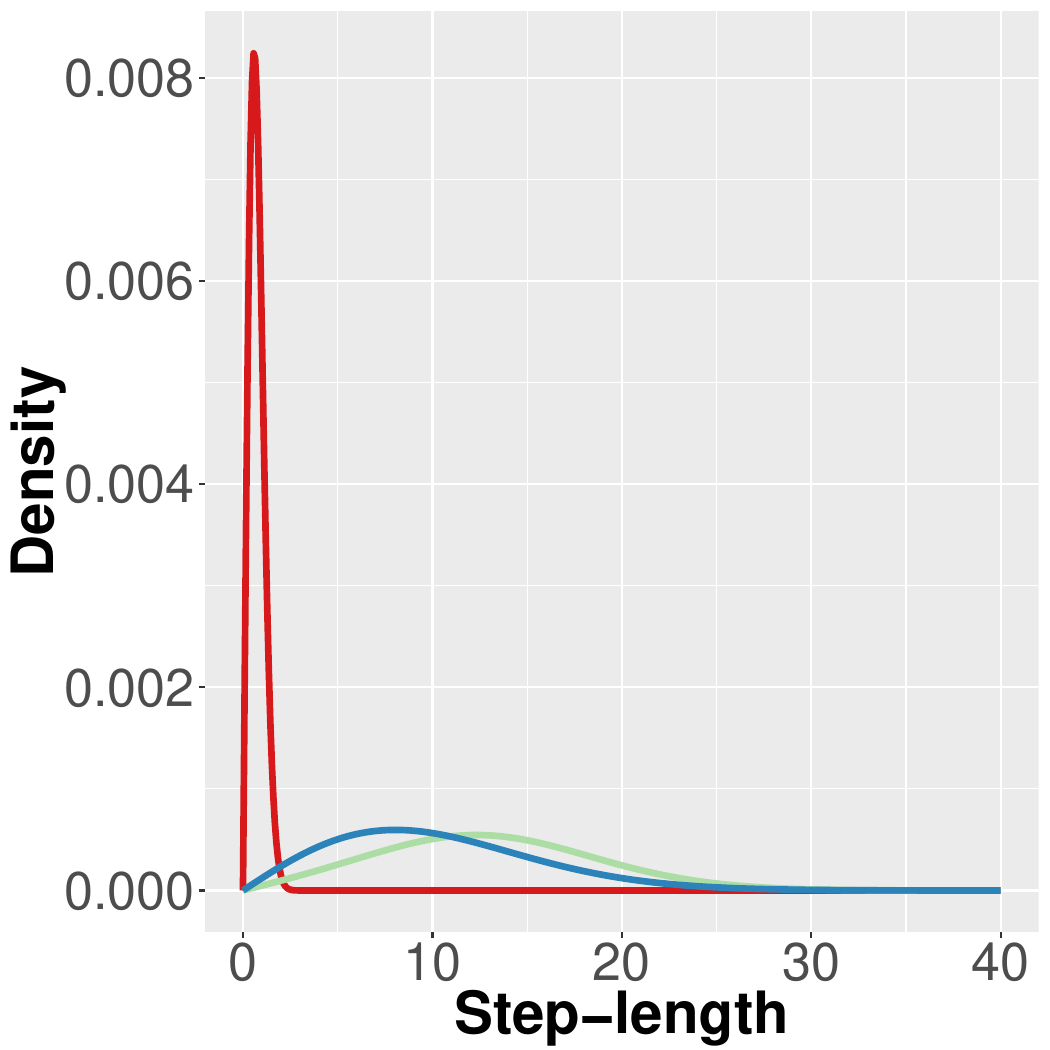}}}
	\caption{The observed spatial locations in the first (a) and second (b) time windows, predictive distributions of  the turning-angle (c) and the step-length (d) for the three different behaviors. The behaviors are encoded with the same colors as in Figure \ref{fig:estTOT}.} \label{fig:DensCL} \label{fig:traj}
\end{figure}

\begin{figure}[t!]
	\centering
	{\subfloat[]{\includegraphics[scale=0.28]{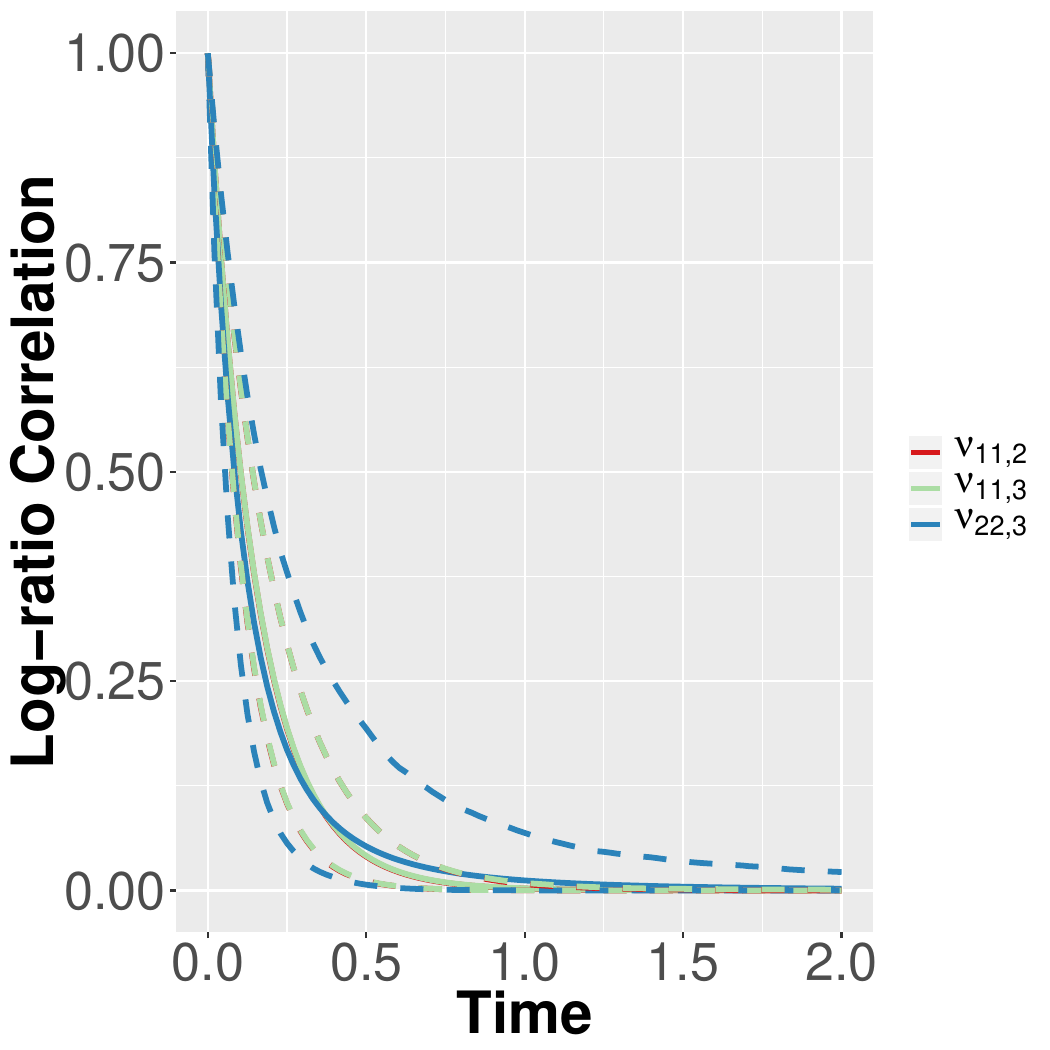}}}
    {\subfloat[]{\includegraphics[scale=0.28]{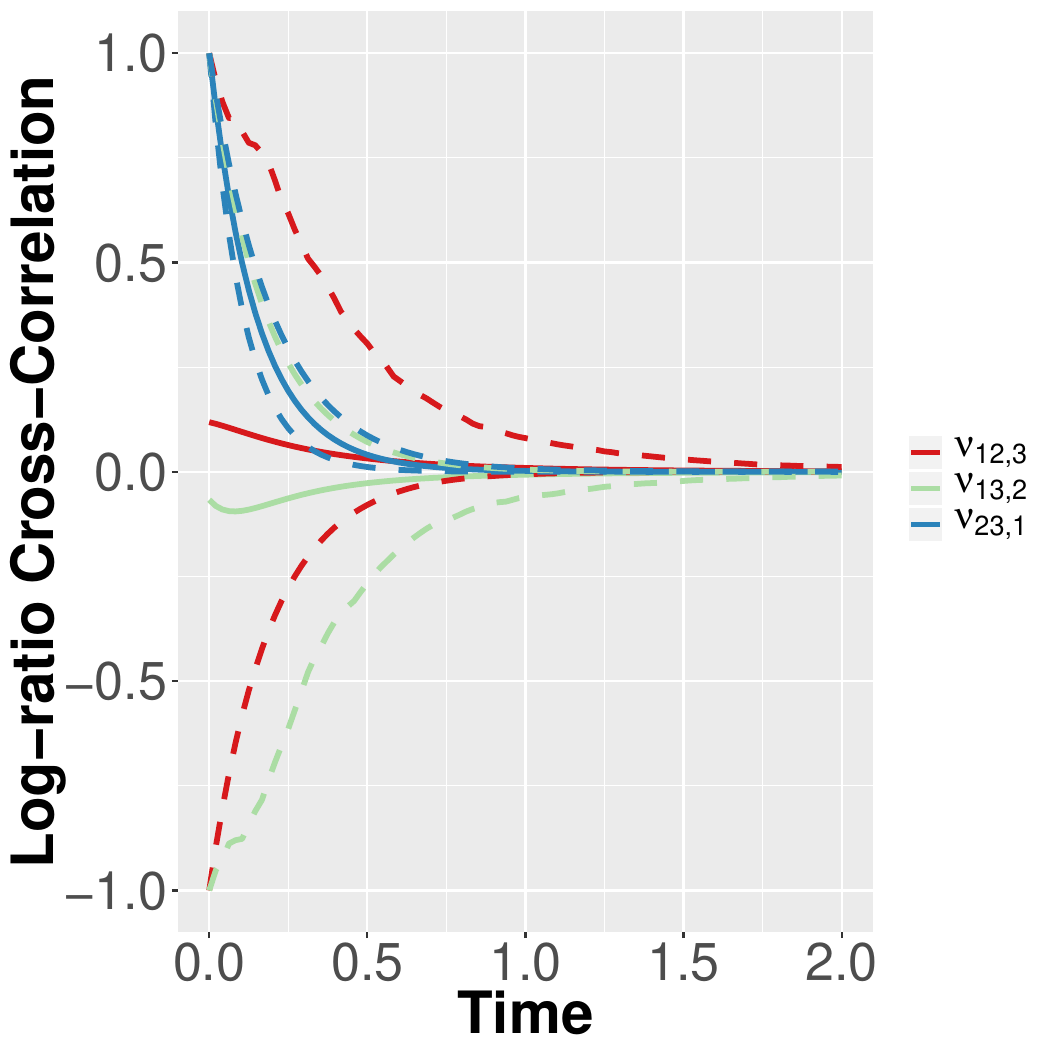}}}
	\caption{Plots of the log-ratios as a function of the temporal distance (expressed in days).   The solid lines represent the posterior means, while the dotted lines are the limits of the 95\% CIs. } \label{fig:logratio}
\end{figure}

\begin{table}[t]
\centering
    \begin{tabular}{l|ccccccc}
        \hline \hline
       Model &$K=2$ & $K=3$& $K=4$& $K=5$& $K=6$\\ \hline
       HMM &                    -12527& -14926                & -13984           &-11941           &-10312 \\ \hline
        CT- HMM             & -13511& -15034                & {-14871}&{-12951}&{-10801} \\ \hline
        LogitGP&     {-13831}  &\textbf{-15925} &-14129             & -12010           & -10210\\
   
\hline
\hline
    \end{tabular}
\caption{ICLs of  the tested models. the best model is in bold. Numbers are rounded to the closest integer.} \label{tab:ICL}
\end{table}

\begin{table}[t]
\centering
    \begin{tabular}{l|ccc}
        \hline \hline
        & $k=1$ & $k=2$ & $k=3$\\ \hline
        $\left[\boldsymbol{\xi}_{k}\right]_1$ &-0.014 & 0.386& -0.543  \\
(CI)&   (-0.017 -0.004)& (0.251 0.531)& (-1.411  0.081) \\ 
        $\left[\boldsymbol{\xi}_{k}\right]_2$   &0.001& 0.061 &0.048   \\
(CI)&   (-0.005  0.008)& (-0.070  0.121)& (-1.291  1.235) \\ 
        $\left[\boldsymbol{\Omega}_{k}\right]_{11}$ &0.003&0.661  &0.38   \\
(CI)&   (0.003 0.004)& (0.533 0.819)& (0.176 0.635) \\ 
        $\left[\boldsymbol{\Omega}_{k}\right]_{12}$ &0.000 & 0.052& 0.001  \\
(CI)&   (0     0)& (-0.044  0.153) & (-0.032  0.030) \\ 
        $\left[\boldsymbol{\Omega}_{k}\right]_{22}$ &0.002 &  0.513&0.381  \\
(CI)&   (0.002 0.004)& (0.403 0.637) & (0.169 0.639) \\ 
$\beta_{k1}$  & 7.395& 4.252&   \\
(CI)&   (6.764  8.94)& (1.848 6.932 ) & ()\\ 
$\beta_{k2}$  & -0.382 &-7.476 &   \\
(CI)&   (-0.879 0.153)& (-11.352 -4.796) & ()\\ 
        $\boldsymbol{\psi}_{k}$ &6.746 &  12.276 & 5.221\\
(CI)&   (4.887 9.187)& (6.135 14.862)& (0.909 11.179) \\ 
$\left[\boldsymbol{\Sigma}\right]_{kk}$ & 143.95&  1.785& 1.731 \\
(CI)&   (40.475 371.745)& (0.221 8.715)& (0.027 11.549) \\ 
\hline \hline 
 &  $\left[\boldsymbol{\Sigma}\right]_{12}$ &  $\left[\boldsymbol{\Sigma}\right]_{13}$ &  $\left[\boldsymbol{\Sigma}\right]_{23}$  \\  \hline
& -4.418 &-5.877 &   1.408 \\
(CI)& (-38.881  19.974) &   (-43.148  13.200)& (0.146  9.337)\\ 
\hline
\hline
    \end{tabular}
\caption{Real data - Posterior means and 95\% CIs of the parameters of the proposed model with $K=3$.} \label{tab:DensCI}
\end{table}

Table \ref{tab:ICL} shows the ICL for all the tested models.  The results of each model suggest the presence of three different behaviors ($K=3$), and our proposal is the one with the lowest ICL (the model with the best goodness-of-fit). 
The posterior estimates, means and 95\% credible intervals (CIs) of the chosen model can be seen in Table \ref{tab:DensCI}, while Figure \ref{fig:estTOT}  shows the posterior estimates of the  probability vector time series. Figure \ref{fig:traj} shows the  observed spatial locations with the associated classification and predictive densities of the step-length and turning-angle. As in the introduction, we indicate the first behavior with R,  the second one with HE1 and  the third one with HR2.
 
\subsection{Behavior description}

\paragraph{First behavior (R)}
From Figure \ref{fig:traj} (c) and (d), it is possible to observe that   the speed  in the first behavior  is very close to zero, and the circular distribution, even though  it has a mode at around $\pi$, has a high variability, thus showing that there is not  a clear preferred direction. The means of the associated GP in both windows (Table \ref{tab:DensCI}) are larger then  the ones of the other behaviors, thus indicating that it is the one with the highest (mean) probability, as it is confirmed in Figure \ref{fig:estTOT}, where it is evident that the probability values are often equal to 1.
This  behavior may be described as a slow-movement behavior, which represents  a variety of activities, such as resting, feeding, social interacting and, in the second window, attending cubs during the reproduction period. It should be noted that the occurrences of this behavior, in the second time window, are spatially localized over a relatively small area, which can reasonable be  identified as the den, see Figure \ref{fig:traj} (b).

\paragraph{Second behavior (HE1)}
As far as the second behavior is concerned, the speed increases and the circular distribution has a well-defined mode around zero, thus indicating that the wolf tends to move in a straight line, see Figure \ref{fig:traj}. This behavior is relevant in the first time window while  it almost disappears in the second, as can be seen from  the associated regressive coefficients. 
This behavior represents the nomadic phase of wolf movement patterns during winter, when the main activities are hunting and patrolling the territory \citep{Mech2003}. Moreover,  F24  established her home range in March 2010, and these high speeds may also represent the  need to control and mark the territory, as newly formed pairs are the ones with the highest marking rates in wolf populations \citep{Rothman1979}.

\paragraph{Third behavior (HE2)}
As far as the third behavior is concerned, the predictive distribution of the step-length is similar to that of the second behavior (Figure \ref{fig:traj} (d)). The turning-angle has a mode at approximately 2 and limited variability, thus indicating that the animal moves in an anticlockwise direction. 
This behavior is almost absent in the first temporal window, while it represents the main movement type in the second one. This pattern seems to correctly estimate  the star-shaped movements of wolves in the presence of cubs at dens \citep{Mech2003}. This is in line with the tendency of breeding females to restrict their movements to a smaller portion of the territory during the period of reproduction, compared to other times of the year   \citep{Jedrzejewski2001}. The counter-clockwise tendency  may be related to the necessity of exploiting different portions of the home range to locate vulnerable prey. Since wolves  have a spatial map of resources within their territory \citep{Peter1979}, varying their hunting routes to surprise prey could improve their hunting success \citep{Jedrzejewski2001}, and this could result  in a rotational use of the home range \citep{Demma2011}.\\

It is possible to see, from the off-diagonal elements of $\boldsymbol{\Sigma}$ (Table \ref{tab:DensCI}), that there could be dependence; it should be recalled that independence between the elements of the compositional vectors requires a diagonal $\boldsymbol{\Sigma}$. To  better analyze the dependence structure we  look at  the temporal evolution of the  ``correlation''. 
 There is no  unique or generally accepted way of evaluating the temporal correlation of compositional data, see, for example,
\cite{Filzmoser2008} or
\cite{LONG2013}, but since we are here mainly interested  in the temporal evolution of the dependence, the log-ratio covariance, equation \eqref{eq:tau1}, is divided by its value when $t= t'$ (at lag 0), thus showing  how dependence changes over  time. These log-ratio  correlations, indicated with $\nu$,  are shown in Figure \ref{fig:logratio}.

It is interesting to note that $\nu_{11,2}$ and $\nu_{11,3}$, being the correlation functions of the second and third behaviors with respect to the first one, are indistinguishable, thus highlighting  that the difference  is mostly due to the direction of movement (the turning-angle). In Figure \ref{fig:logratio} (b), which shows the cross-correlations, only  $\nu_{23,1}$   has 95\% CIs  that does not contain  zero. 
Moreover, dependence almost disappears for a lag of 12 hours (all the correlation values are close to zero) and it is numerically zero after 24 hours. From a biological point of view, this means that the probability of following a particular behavior at time $t$ is  influenced to a great extend by the probabilities and behaviors  of the 12 previous hours, is  slightly influenced  by the  12 to 24 previous hours, but what the animal did the day before shows no influence.

\subsection{Time window description}

\paragraph{First time window}
In the first window, the  slow-movement  behavior (R) has a high probability of occurring during daylight hours, whereas the HR1 seems to be more likely during the night (Figure \ref{fig:estTOT} (a)).
This complementary pattern is in line with the circadian activity of wolves in human-modified environments, where they are mainly nocturnal to avoid disturbances  from human activities during the day  \citep{ciucci1997,Theuerkauf2009}.

\paragraph{Second time window}
In the second window, the slow-movement regime has a probability close to one during the first days, because  F24 likely entered the den and gave birth during that time. According to previous research on wolf reproducing behavior, breeding females are stationary on the day of birth, and with limited movements during the following period \citep{Alfredeen2006}. During the days after reproduction occurred, the slow-movement regime is often concentrated around dusk  and during the first hours of the night. Instead, the star-shaped moving regime is concentrated during daylight hours, and shows two or more peaks at different times of the day (Figure \ref{fig:estTOT} (b)). This result can be interpreted as a reduction in the nocturnal activity of this wolf due to the presence of cubs, which is also accompanied by a relative increase in diurnality. During the reproduction period, breeding females spend most  of the time at their den and rendezvous sites  \citep{Ballard1991,Harrington1982}. Since other wolves from the pack usually ensure the feeding of breeding females during this time \citep{Mech1999}, females do not have to maintain an activity pattern based on hunting which, in our study area, may be nocturnal due to  the  presence of humans. 
This situation may have led  F24 to mainly leave the den  during the day, when sunlight can help  keep the unattended cubs warm, and other large carnivores (such as Apennine brown bears) are less active \citep{Vila1995}.

It should be noted that, since the temporal evolution of $\boldsymbol{\pi}$ is modeled non-parametrically, we found the  daily-temporal patterns in Figure \ref{fig:estTOT} without introducing any  constraint into our model or introducing factors that can be used to model it directly.

\section{Final remark}
\label{sec:conclu}

Motivated by our dataset, in which a female wolf is observed over two time windows,
we  have proposed a novel approach to analyze  trajectory tracking data. This approach is aimed at defining the posterior distribution of the clustering probabilities, where the clusters are representative of different behaviors that the animal exhibits, and  at describing the trajectory conditionally on the particular behavior by characterizing the associated step-length and  turning-angle.

Our model is based on a  Bayesian non-parametric approach, where the time varying probabilities $\boldsymbol{\pi}(t)$  are modeled through a  $\text{LogitGP}$. Our model is invariant with respect to the arbitrary choices of the reference element and the reordering of the labels.  We have  proposed fitting the model under a Bayesian framework and the number of latent behaviors through ICL, which is one of the most frequently used information criterion.

We  fitted the model on the wolf data.  The results we obtained are easy to interpret and give  insight into  wolf behavior, both in terms of movement metrics (in particular, step-length and turning-angle) and the time evolution. 

The advantage of a non-parametric approach is that even though some features of the data are not modeled explicitly, posterior inference may reveal them afterwards. An example is the attractive effect of the unknown location of the den. The distribution of the turning-angle in  HE2  has a peak close to  $\pi$, which may be  in fact interpretable as the effect of such an attraction \citep{Parton2017}.
A second example is the seasonality of the components of $\boldsymbol{\pi}(t)$  which is discussed in detail in Section \ref{sec:realdata}.

The model we have proposed was motivated by a trajectory tracking dataset, but it could be employed in different contexts, in particular in the case of environmental sciences, where spatial information can be incorporated in the  probability dependence structure, so that the response variable behaves in a similar way in  locations that are close in space. This aspect will be investigated   in the future.
 
Another issue that deserves further attention is the estimation of the unknown number of clusters. For the moment, we decided to fit the model with different numbers of components, and then use  a model-fit measure to select the \emph{best} one. A fully Bayesian analysis, where a prior distribution is defined for the number of components, would also be very interesting.

Future lines of research include the extension of the present model to multi-animal trajectories. The generalization of the model proposed in this work would be an extension of the hidden Markov models proposed in \cite{langrock2014modelling}, with relation to the continuous-time approach of  \cite{niu2016modeling}.

\section{Acknowledgments}
We acknowledge the support of the Italian Ministry of Education, University and Research (MIUR) grant \emph{Dipartimenti di Eccellenza}, CUP: E11G18000350001, conferred to the Dipartimento di Scienze Matematiche - DISMA, Politecnico di Torino.
The work of the first two authors has partially been developed under the PRIN2015 supported-project Environmental processes and human
activities: capturing their interactions via statistical methods (EPHAStat), funded by MIUR (Italian Ministry of Education, University and
Scientific Research) (20154X8K23-SH3). 
 
The GPS  data  were collected within the ``Large Carnivores'' project, which took place from 2006 to 2011 in the Abruzzo, Lazio and Molise National Park. The project was supervised by the Department of Biology and Biotechnology ``Charles Darwin'' of the University of Rome ``La Sapienza'' and funded by a private American donor through the ``Wildlife Conservation Society''.
 
The authors would like to thank Tilman Gneiting, the editor, the associate editor and the referees for their  valuable suggestions.

\bibliographystyle{ba}
\bibliography{all}

\appendix
  
\section{MCMC implementation} \label{sec:impl}

\begin{figure}[t]
	\centering
	{\subfloat[]{\includegraphics[scale=0.25]{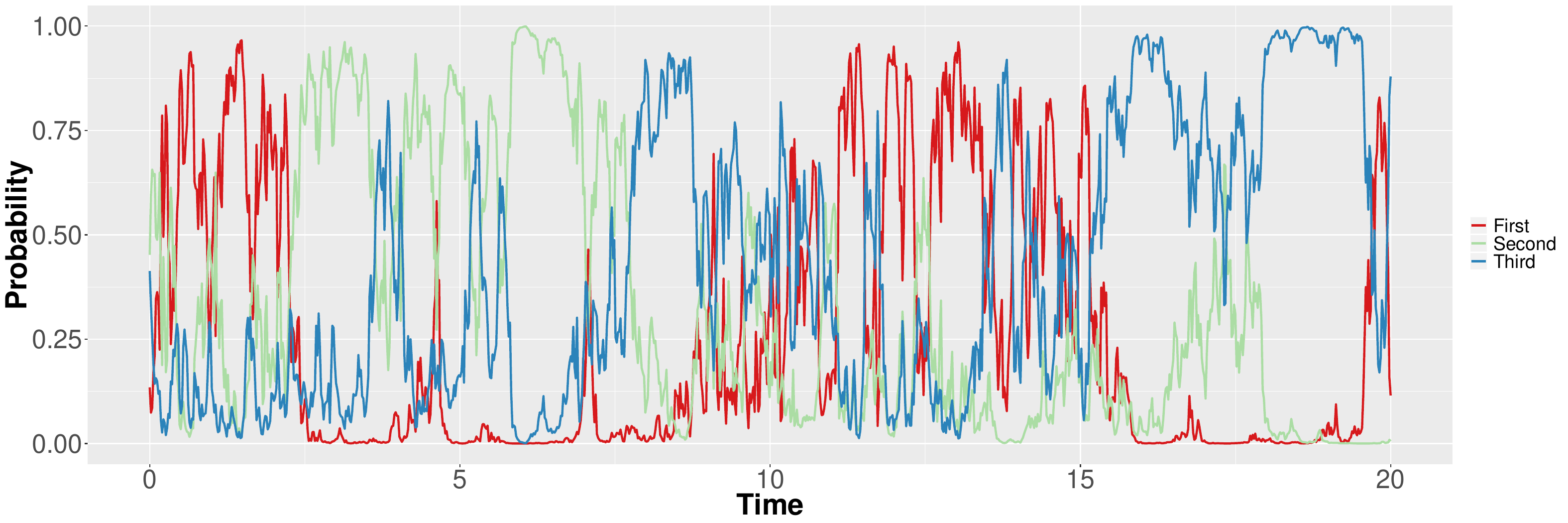}}}\\
	{\subfloat[]{\includegraphics[scale=0.25]{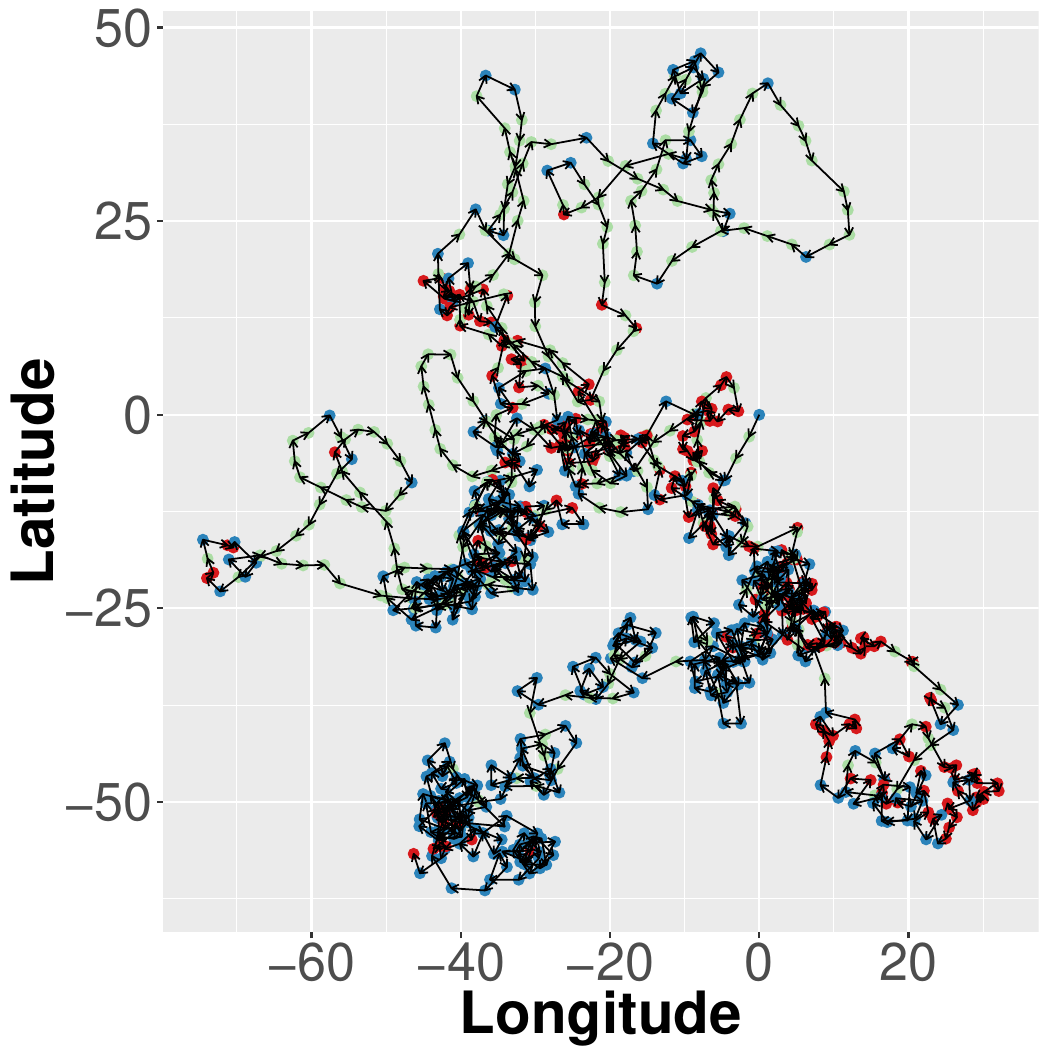}}}
	{\subfloat[]{\includegraphics[scale=0.25]{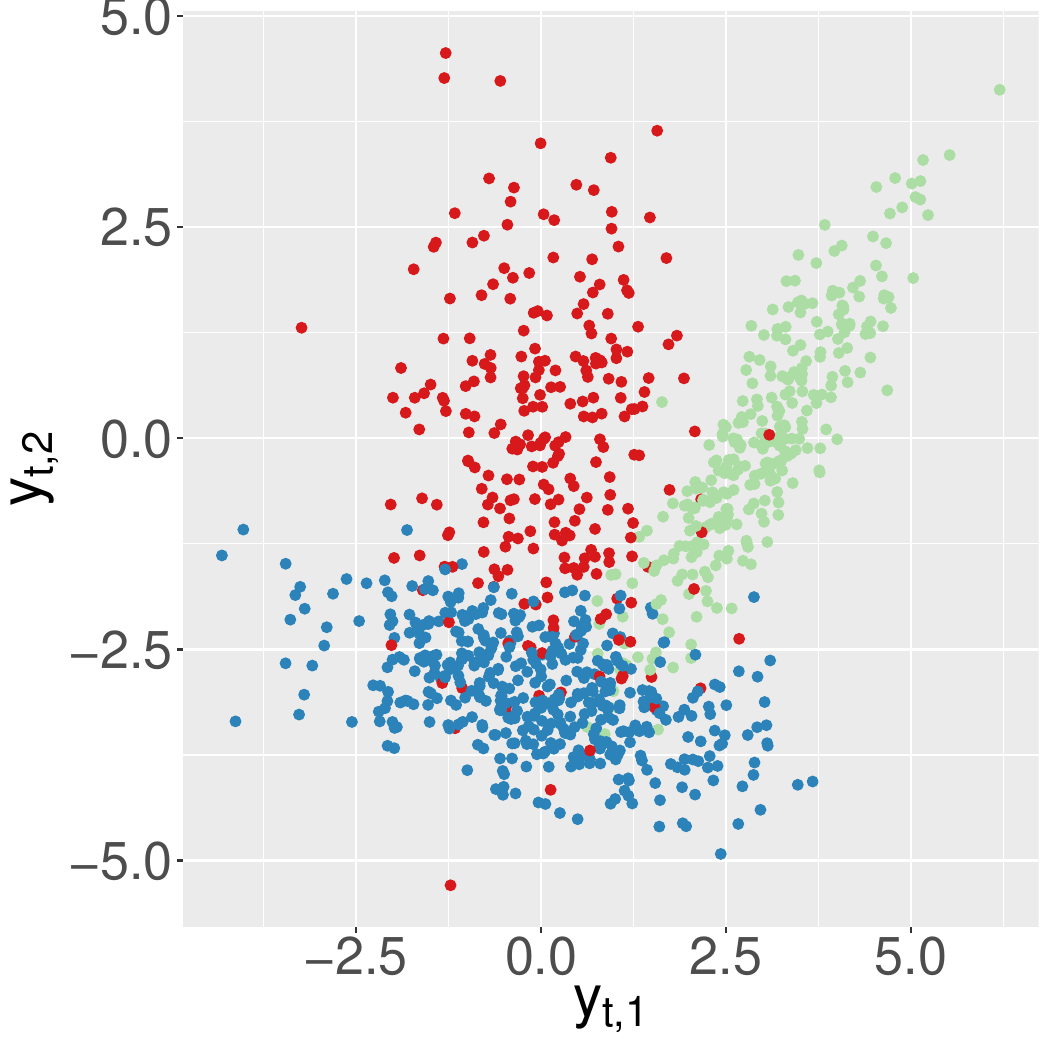}}}
	\caption{Simulated example -  Probability vector time series (a), trajectory (b), data $\mathbf{y}$ (c). The colors represent the  different behaviors.} \label{fig:estb3}
\end{figure}

\begin{figure}[t]
	\centering
	\includegraphics[scale=0.25]{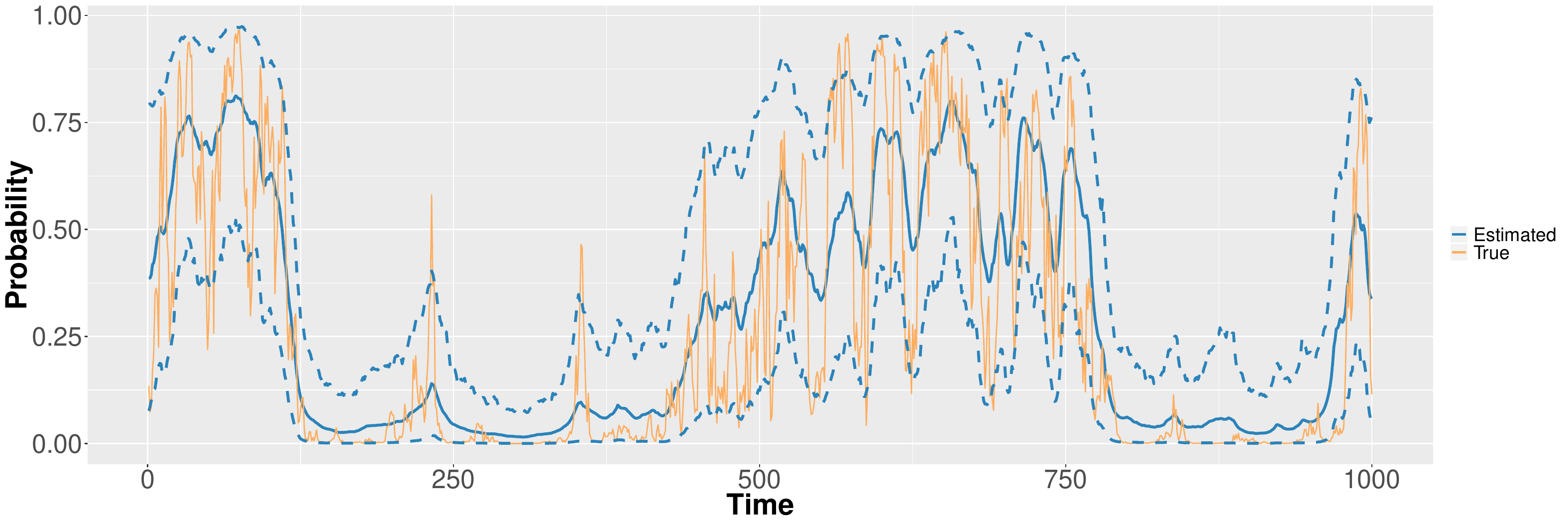}
	\caption{Simulated example - Posterior estimates of the probability vector time series of the first behavior. The solid lines represent the posterior means while the dotted lines are the limits of the  95\% CIs.} \label{fig:est}
\end{figure}

\begin{table}[t]
	\centering
	\tiny
	\begin{tabular}{l|ccc}
		\hline \hline
		&  $k=1$ & $k=2$ & $k=3$ \\ \hline
		$\left[\boldsymbol{\xi}_{k}\right]_1$ & 0.043&2.953 &-0.234 \\
		(CI)&   (-0.165  0.277)& (2.745 3.236) & (-0.441 -0.112)\\
		$\left[\boldsymbol{\xi}_{k}\right]_2$  & -0.098&-0.124 & -2.897 \\
		(CI)& (-0.467  0.301)& (-0.364  0.117) & (-3.001 -2.791)\\
		$\left[\boldsymbol{\Omega}_{k}\right]_{11}$ &  1.446&1.271 & 2.076 \\
		(CI)&(1.074 1.886)& (1.015 1.625) & (1.764 2.427)\\
		$\left[\boldsymbol{\Omega}_{k}\right]_{12}$ & -0.004& 1.464& -0.539 \\
		(CI)&(-0.373  0.356)& (1.151  1.841) & (-0.706 -0.389)\\
		$\left[\boldsymbol{\Omega}_{k}\right]_{22}$ &3.338&2.351 &  0.635  \\
		(CI)&   (2.518 4.182)& (1.887 2.916) & (0.519 0.771)\\
		$\beta_{k0}$ &0.580 & 2.206 & \\
		(CI)&    (-1.955  3.169)& (0.764 3.721) & ()\\
		$\beta_{k1}$  &-2.508 &-5.09 &  \\
		(CI)&    (-6.743  1.463)& (-7.724 -2.649) & ()\\
		$\boldsymbol{\psi}_{k}$ &0.987 &2.247 &  2.89\\
		(CI)&   (0.563 1.639)& (0.422 5.361) & (0.652 5.208)\\
		$\left[\boldsymbol{\Sigma}^*\right]_{kk}$& 3.889&0.78 & 1.979\\
		(CI)&   (0.358 8.784)& (0.143 3.579) & (0.549 5.001)\\
		\hline \hline 
		&  $\left[\boldsymbol{\Sigma}\right]_{12}$ &  $\left[\boldsymbol{\Sigma}\right]_{13}$ &  $\left[\boldsymbol{\Sigma}\right]_{23}$ \\  \hline
		&  -0.118& 0.557&  0.338 \\
		(CI)&   (-1.072  1.479)& (-1.292  3.684) & (-0.891  3.150)\\ 
		\hline
		\hline
	\end{tabular}
	\caption{Simulated Example -  Posterior means and 95\% CIs under $K=3$.} \label{tab:res1}
\end{table}
The MCMC implementation is straightforward. Given a value  of the entire  multivariate GP, its parameters can be simulated as in the usual GP framework. In details, we update all the parameters  at the same time using a  Metropolis step, according to algorithm 4 of \cite{andrieu2008}. Before applying the  algorithm, it is necessary to transform  the parameters so that they   belong to $\mathbb{R}$. Then, we  take the logarithm of the decay parameters and in order  to eliminate the constraints over the parameters of the non-negative definite matrix $\boldsymbol{\Sigma}$,  we re-express it using the Bartlett decomposition   \citep{anderson2003introduction}, which is based on  random variables that are normally and chi-squared distributed; the latter is then transformed using the logarithm. 
Given the probabilities $\boldsymbol{\pi}$ and the data $\mathbf{y}$, the parameters and the latent variable $\mathbf{z}$ are simulated as in a mixture model using Gibbs steps. In order to simulate the GP elements, we use the novel approach of  \cite{polson2013} and its extension, as proposed in \cite{Linderman2015}.

When the missing $\mathbf{y}$ are simulated, we have to ensure that their values are ``coherent'',   meaning that $\mathbf{y}$ have to define a trajectory $\mathbf{s}$ that goes through   the observed locations. 
Equation \eqref{eq:traj} shows that  having a realization of $\mathbf{s}$  we can easily compute  $\mathbf{y}$. However, it is easier to simulate $\mathbf{s}$ since the animal trajectory follows an ARMA-type model.
Therefore,
the missing observations are obtained by first simulating the missing $\mathbf{s}$  and then deriving the corresponding $\mathbf{y}$; in this way, the uncertainty implicit in the presence of missing data is automatically incorporated in the inferential method.

All codes are available from the first author, upon request.

\section{Simulated examples}
\label{sec:simstud}

In this section, we show that the MCMC is able to estimate the  parameters in a satisfactory way and, moreover, that  ICL retrieves  the right number of latent classes.

We  simulate data with $K=3$, and three different  number of observations $T=250,500,1000$ over a time window of length 20. The time-lag between observations is assumed constant  and therefore $t_{i}-t_{t-1}=20/T$; in other words, the intervals between the observations decrease as $T$ increases. The parameters are $\boldsymbol{\xi}_1 = (0,0)'$, $\boldsymbol{\xi}_2 = (3,0)'$, $\boldsymbol{\xi}_3 = (0,-3)'$ and 

\begin{equation}
\boldsymbol{\Omega}_1= \left( 
\begin{array}{cc}
1&0\\
0&3
\end{array}
\right),\,
\boldsymbol{\Omega}_2= \left( 
\begin{array}{cc}
1&1.272\\
1.272&2
\end{array}
\right),\,
\boldsymbol{\Omega}_3= \left( 
\begin{array}{cc}
2&-0.5\\
-0.5&0.5
\end{array}
\right).
\end{equation}

We also assume
$$
\boldsymbol{\Sigma}=
\left( 
\begin{array}{ccc}
5 &  -2 &   0\\
-2 &   5   & 3\\
0  &  3  &  5
\end{array}
\right),
$$
and use exponential correlation functions with decay parameters equal to $1$, $0.8$ and $1.5$, respectively. 
We assume $\mu_1(t)= \beta_{10}+\beta_{11}t $ and $\mu_1(t)= \beta_{20}+\beta_{21}t $ with    $\beta_{10}=0$, $\beta_{11}=-5$, $\beta_{20}=3$, $\beta_{21}=-7$ for the mean function of the GPs.

We simulate 100 datasets for each $T$,  and  we perform inference by fixing $K$ to values between 2 and 6. To ensure that  three separate clusters can be identified, datasets with a  cluster composed of less than  5\% of the total sample size are discarded.  Model choice is performed using the ICL.

We assume  $\boldsymbol{\xi}_k \sim N_2(\mathbf{0}_2, 100\mathbf{I}_2)$ and $\boldsymbol{\Omega}_k \sim IW(3, \mathbf{I}_2)$ as  prior distributions  for the likelihood parameters, $U(0.3,6)$ for the temporal decays,  the regression coefficients are normally distributed with a mean  of $0$ and a variance of $100$ while $\boldsymbol{\Sigma} \sim IW (K+1, \mathbf{I}_K)$.  The MCMC is implemented with 1,000,000 iterations, burnin 70,000 and thin 6, which results in  5,000 posterior samples.

The ICL selects  $K=3$  95\% of the times with $T=250$, $99\%$ with $T=500$ and $100\%$ if $T=1000$.
In order to obtain a better insight into this result, we randomly selected  one dataset with   $T=1,000$, and we show the parameter estimates  in Table \ref{tab:res1}, while the posterior estimates of the compositional vector  time series, with the  associated 95\% CIs,  are depicted,  in Figure \ref{fig:est},  for the first behavior. The ``true'' compositional vector time series, trajectory and $\mathbf{y}$ are shown in Figure \ref{fig:estb3}. From these it is possible to see that most of the true values of the parameters are inside the associated  95\% CIs, as well as  the true compositional  vector time series.

\end{document}